\documentclass[10pt]{article}
\usepackage{graphicx}
\usepackage{amsmath,amssymb}
\usepackage{multirow}
\usepackage{array}
\usepackage{url}
\usepackage{enumerate}
\usepackage{cite}
\usepackage{color} 
\usepackage{dcolumn}
\usepackage{floatrow}
\usepackage{subcaption}
\usepackage{longtable}
\usepackage[labelfont=bf,labelsep=period,justification=raggedright]{caption}
\makeatletter
\renewcommand{\@biblabel}[1]{\quad#1.}
\makeatother
\date{}

\newfloatcommand{capbtabbox}{table}[][\FBwidth]

\begin{document}

\begin{flushleft}
{\Large
\textbf{Exploring the Uncharted Export: an Analysis of Tourism-Related Foreign Expenditure with International Spend Data}
}
\\ 
 \bigskip
Michele Coscia$^{1,*}$, Ricardo Hausmann$^{1}$ and Frank Neffke$^{1}$
\\
\medskip
\bf{1} Center for International Development, Harvard University, Cambridge MA, USA
\\
\smallskip
$\ast$ E-mail: Corresponding Author michele\_coscia@hks.harvard.edu
\end{flushleft}

\section*{Abstract}
Tourism is one of the most important economic activities in the world: for many countries it represents the single largest product in their export basket. However, it is a product difficult to chart: ``exporters'' of tourism do not ship it abroad, but they welcome importers inside the country. Current research uses social accounting matrices and general equilibrium models, but the standard industry classifications they use make it hard to identify which domestic industries cater to foreign visitors. In this paper, we make use of open source data and of anonymized and aggregated transaction data giving us insights about the spend behavior of foreigners inside two countries, Colombia and the Netherlands, to inform our research. With this data, we are able to describe what constitutes the tourism sector, and to map the most attractive destinations for visitors. In particular, we find that countries might observe different geographical tourists' patterns -- concentration versus decentralization --; we show the importance of distance, a country's reported wealth and cultural affnity in informing tourism; and we show the potential of combining open source data and anonymized and aggregated transaction data on foreign spend patterns in gaining insight as to the evolution of tourism from one year to another.

\section{Introduction}
Tourism is an atypical export sector. Instead of shipping products outside a country, the exports -- mostly locally consumed services -- never leave the country. Instead, the importers themselves -- tourists -- travel to the country to make their purchases. Given this peculiar structure, it is not easy to quantify the impact of the tourism sector in the economy of a country, even if it is considered one of the most important economic activities in the world \cite{stabler2009economics}. Currently, there are a variety of techniques to estimate the size of the tourism sector in a country. The literature ranges from modeling external capital flows with social accounting matrices (SAM) and computable general equilibrium models \cite{zhou1997estimating}, to surveys from a sample of participants \cite{kaylen1998estimating, frechtling2000assessing} -- i.e. ``primary surveys'' -- or from employment compensations that are hypothesized to be related with tourism activities \cite{leatherman1996estimating} -- ``secondary surveys''. Combinations of both SAM models and surveys are also used \cite{daniels2004estimating}. Researchers also use special occasions to evaluate the impact of tourism, typically large and recurring sport events \cite{hodur2007estimating}, but also environmental quality changes \cite{turpie2001estimating} and disasters \cite{garza2006estimating}.

None of these methods is perfect. Issues are usually grouped into four categories: substantive issues, aggregation issues, structural issues of change and prediction, and intangible impacts \cite{briassoulis1991methodological}. Among these issues, we are particularly interested in the last one: intangible impacts. To estimate the actual impact of tourism is difficult, because all these models and surveys are not a direct observation of tourism. Surveys shed light on self-reported expenditures by tourists. The SAM and related models also have difficulty in collecting actual tourism data: one can easily identify hotels and attractions as being part of the tourism sector, but this would ignore the fact that tourists also spend money elsewhere. A tourist might dine at a restaurant that is not considered part of the tourist sector because it is located in a neighborhood mostly visited by local residents. The same holds for many other activities: concerts, spas and events, but also retail and medical expenses should be counted as exports. Direct observation data is usually limited to very specific sectors and situations, like national parks \cite{eagles2000estimating}.

In this paper, we propose a methodology to address this issue. We use anonymized and aggregated foreign transaction data provided by the Mastercard Center for Inclusive Growth. A payment card expenditure is foreign if it took place in a country different than the one where the bank issued the card.

Spend patterns enhance the current literature in multiple ways. First, it is a direct observation of foreign demand, skipping the modeling step and providing a direct quantification of these flows. Second, we can utilize metadata about the merchants involved in the transactions with the tourist segment. When registering its point of sale system, the merchant has to specify its address. This enables us to understand the locations that are most popular among foreign travelers, at a level of detail that overall country-wide or state-wide SAMs cannot. Third, there is additional metadata we can analyze regarding the merchant category code. We use an internal merchant classification code to distinguish among different spend categories. We are then able to quantify the impact of sectors that are traditionally not considered as a part of tourism.

\begin{figure}
\centering
\includegraphics[width=.495\columnwidth]{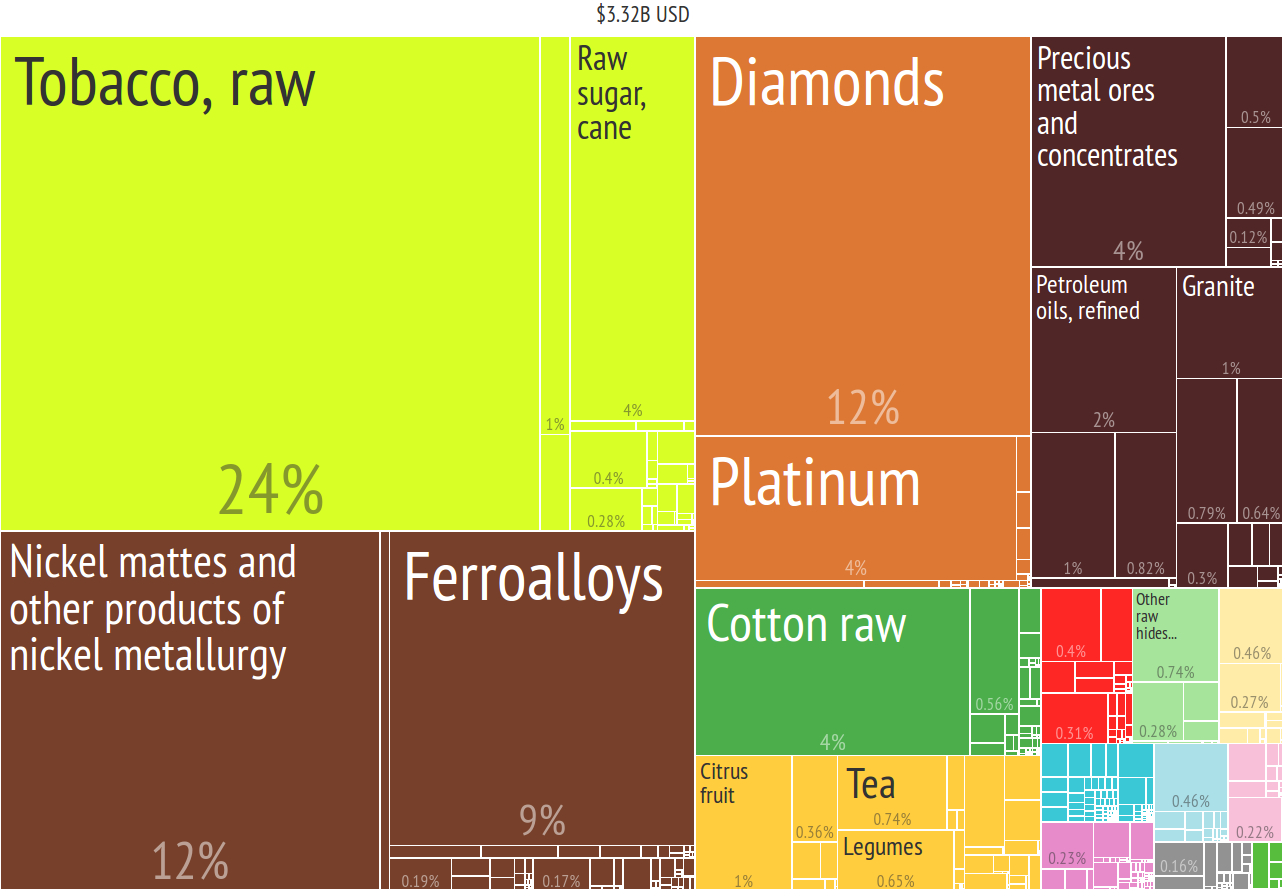}
\includegraphics[width=.495\columnwidth]{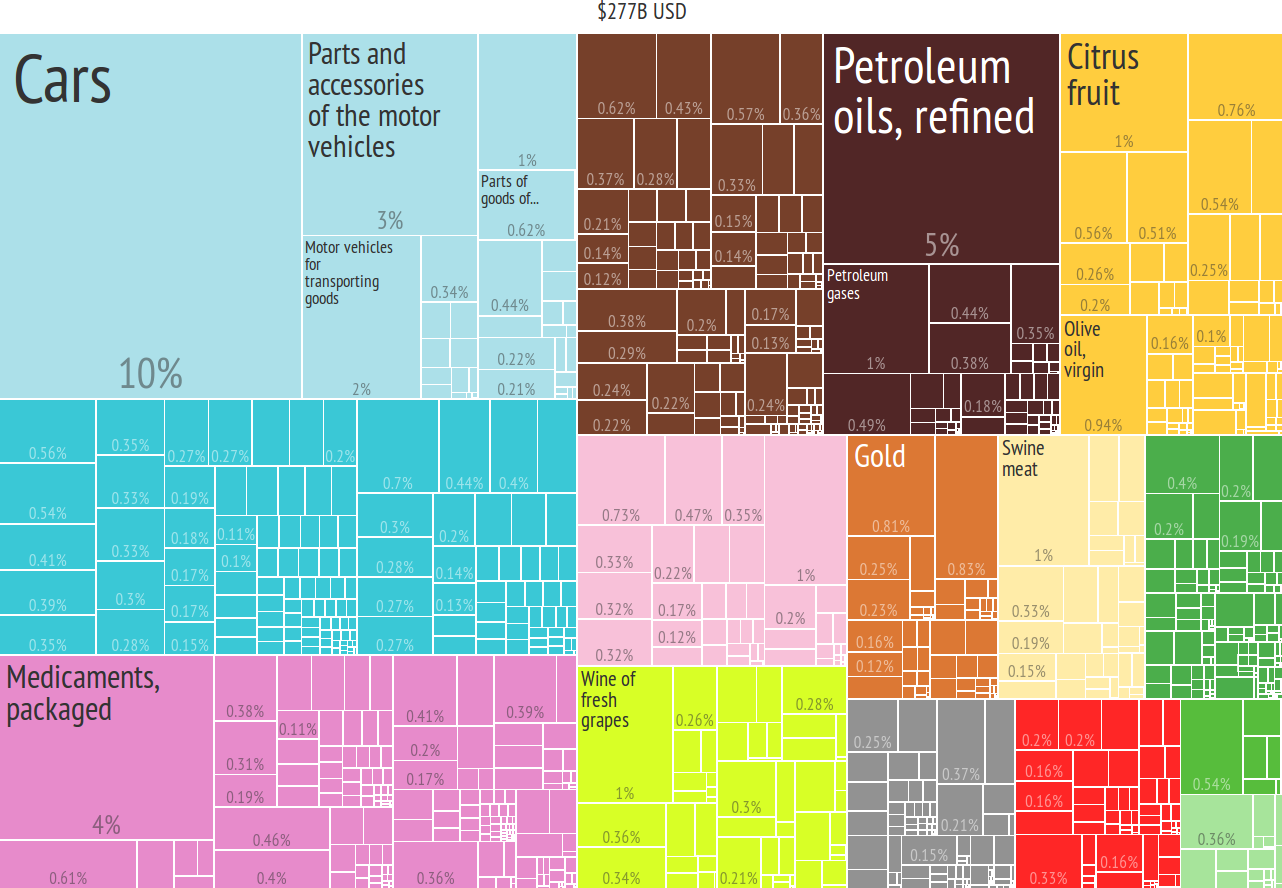}
\caption{The export baskets of Zimbabwe (left) and Spain (right) in 2013. The total gross export amount is reported on top. Color represents related industries in the HS4 classification. Data from the Atlas of Economic Complexity \cite{hausmann2014atlas}, gathered from {\tt http://atlas.cid.harvard.edu/} (date of access: March 9th, 2016).}
\label{fig:export-shares}
\end{figure}

The economic literature agrees that tourism is an important sector in driving economic development \cite{gunduz2005tourism, sinclair1998tourism} and it positively influences household wages \cite{daniels2004estimating}. In some cases, there is no need for sophisticated arguments to show the impact of tourism. Take Zimbabwe as an example. The World Bank estimated receipts of international tourism\footnote{From \url{http://data.worldbank.org/indicator/ST.INT.RCPT.CD} (date of access: March 9th, 2016).} for Zimbabwe in 2013 at 846 million USD. This estimate is imprecise for the reason exposed before but, following our previous arguments, it might be an underestimation\footnote{The World Bank estimates are around five times our estimates, however we explain in the next section why our reported dollar amounts are just a fraction of the actual tourist expenditures.}. In the same year, Zimbabwe exported goods for a total of 3.32 billion USD, as reported in Figure \ref{fig:export-shares} (left). The tourism sector is around a quarter of the total export basket. It is a source of external income larger than any product Zimbabwe is currently exporting. It is larger than tobacco, and larger than diamonds and nickel combined -- the second and third largest export products, respectively. This argument does not hold only for developing economies. It holds even more strongly for developed countries. Consider Spain: with 67 billion USD coming from tourism and a 277 billion USD export basket (Figure \ref{fig:export-shares}, right), tourism represents a similar share of external income -- around 24\%. The most important export of Spain, cars, only totals 28 billion USD, less than half of tourism receipt.

In the paper we show the potential insights to be gained from analyzing foreign payment card expenditures. We focus on two countries for which we have detailed geographical information: Colombia and the Netherlands. We firstly provide some descriptive statistics about the most popular destinations and industries, and the origins that are most attracted to these two countries. We then detail a possible framework that can be used by countries and regions to diagnose the health of their tourism sector. With this tool, locations might be able to identify new opportunities in attracting tourists similar to the ones they are already attracting, or to develop new tourism services. A growing tourism income could be used to foster inclusive economic development, by introducing currently excluded workers into the new productive activities. Finally, we provide a rough estimate of the ``hidden'' tourism sectors: how much do tourists spend on industries that are traditionally not considered part of the sector itself.

\section{Describing Tourism}
In this section we present an overview on the anonymized and aggregated foreign card transaction data in Colombia and the Netherlands. There are two issues with the data. First, we cannot observe imported cash, which anyway is becoming less important nowadays. Second, we need to scale for other payment tenders. Unfortunately, we are unable to perform the country-specific corrections our estimates need for lack of data. These two issues are bound to introduce a certain amount of error. 


The anonymized and aggregated transaction data is derived from October 2011 to September 2014. During this period, the Netherlands received a significantly higher number of foreign cards. This number might be inflated by the large amount of foreigners living near the country, who can freely travel inside it thanks to good integration in the European Union. The total foreign card expenditures in the Netherlands is more than five times higher than Colombia's. However, correcting for purchase power parity the ratio goes down to less than three times, as the cost of living in Colombia is lower: each dollar buys tourists more in Colombia than in the Netherlands. Looking at the number of transactions, it appears that the average number of transactions per traveler in Colombia is higher. This suggests that tourists may visit the country for a longer period, while more of the flow in the Netherlands is composed by day travelers and commuters.

\subsection{Where do tourists go?}
\begin{figure}[!t]
\begin{floatrow}
\ffigbox{
   \includegraphics[width=\columnwidth]{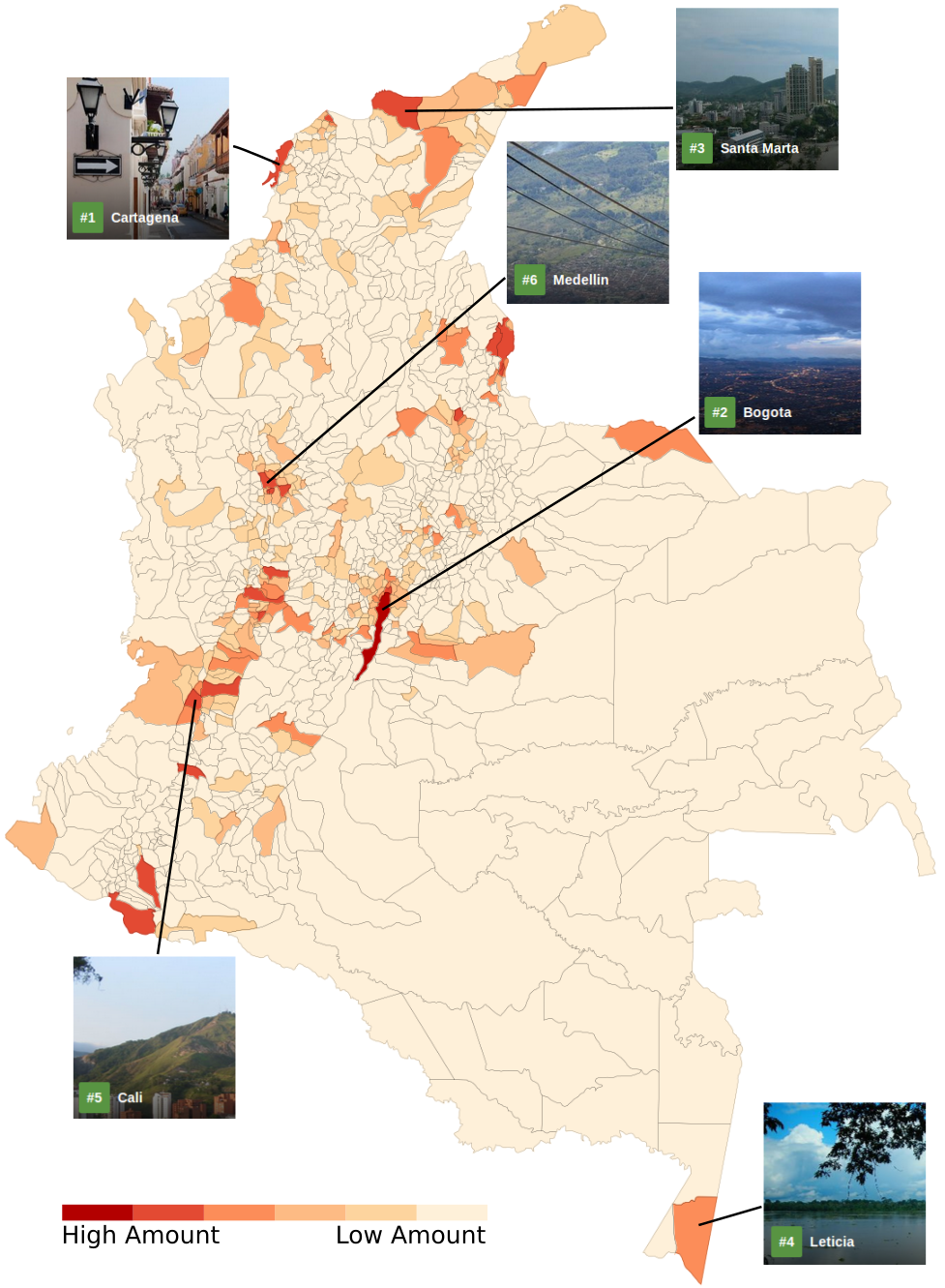}
}{
  \caption{The most popular destinations in Colombia (red = highest USD expenditure, white = low/no USD expenditure). Data corresponds to Table \ref{tab:col-destinations}, color map is in logs. We highlight the top 6 tourism destinations according to Trip Advisor. Aggregated and anonymized data from MasterCard's Center for Inclusive Growth.}
  \label{fig:col-destinations}
}
\capbtabbox{
\begin{tabular}{l|l|rr}
Rank & Zone & \%\\
\hline
1 & Santafe De Bogota D.C. & 39.00\%\\
2 & Cucuta & 18.21\%\\
3 & Cartagena & 9.56\%\\
4 & Medellin & 8.30\%\\
5 & Cali & 6.04\%\\
6 & Barranquilla & 2.53\%\\
7 & San Andres & 1.82\%\\
8 & Santa Marta & 1.48\%\\
9 & Pereira & 1.23\%\\
10 & Envigado & 0.81\%\\
11 & Rionegro & 0.79\%\\
12 & Bucaramanga & 0.79\%\\
13 & Ipiales & 0.62\%\\
14 & Chia & 0.62\%\\
15 & Armenia & 0.58\%\\
16 & Manizales & 0.45\%\\
17 & Palmira & 0.38\%\\
18 & El Zulia & 0.36\%\\
19 & Los Patios & 0.32\%\\
20 & Sabaneta & 0.32\%\\
21 & Popayan & 0.30\%\\
22 & Pasto & 0.30\%\\
23 & Leticia & 0.28\%\\
24 & Ibague & 0.26\%\\
25 & Itagui & 0.23\%\\
\end{tabular}
}{
\caption{The top tourist destinations for Colombia in percentage of dollars spent inside the country across all foreign countries of origin, industries and time periods. Mastercard insights.}
\label{tab:col-destinations}
}
\end{floatrow}
\end{figure}

Figure \ref{fig:col-destinations} and Table \ref{tab:col-destinations} show the geographical distribution of tourist expenditures in Colombia. Large areas of the country are mostly ignored by foreign visitors. This includes the South-East part of the country, which is mostly covered by the Amazon forest. Some centers of aggregation emerge. They can be divided into three classes. The first is the largest cities: Bogota, Medellin and Cali. The second is big attractors of leisure trips: Cartagena is one of the most popular destinations in Latin America.

The third class is border cities: Cucuta (at the border with Venezuela), Ipiales (Ecuador) and Leticia (Brazil and Peru). This last class shows that Colombia still has some important flows of foreign commuters, although not as much as the Netherlands. It also shows the impact of Venezuela on the demand for goods in Cucuta, a phenomenon that is temporary in nature. In September 2015, Venezuela closed the border. 

\begin{figure}[!t]
\begin{floatrow}
\ffigbox{
   \includegraphics[width=\columnwidth]{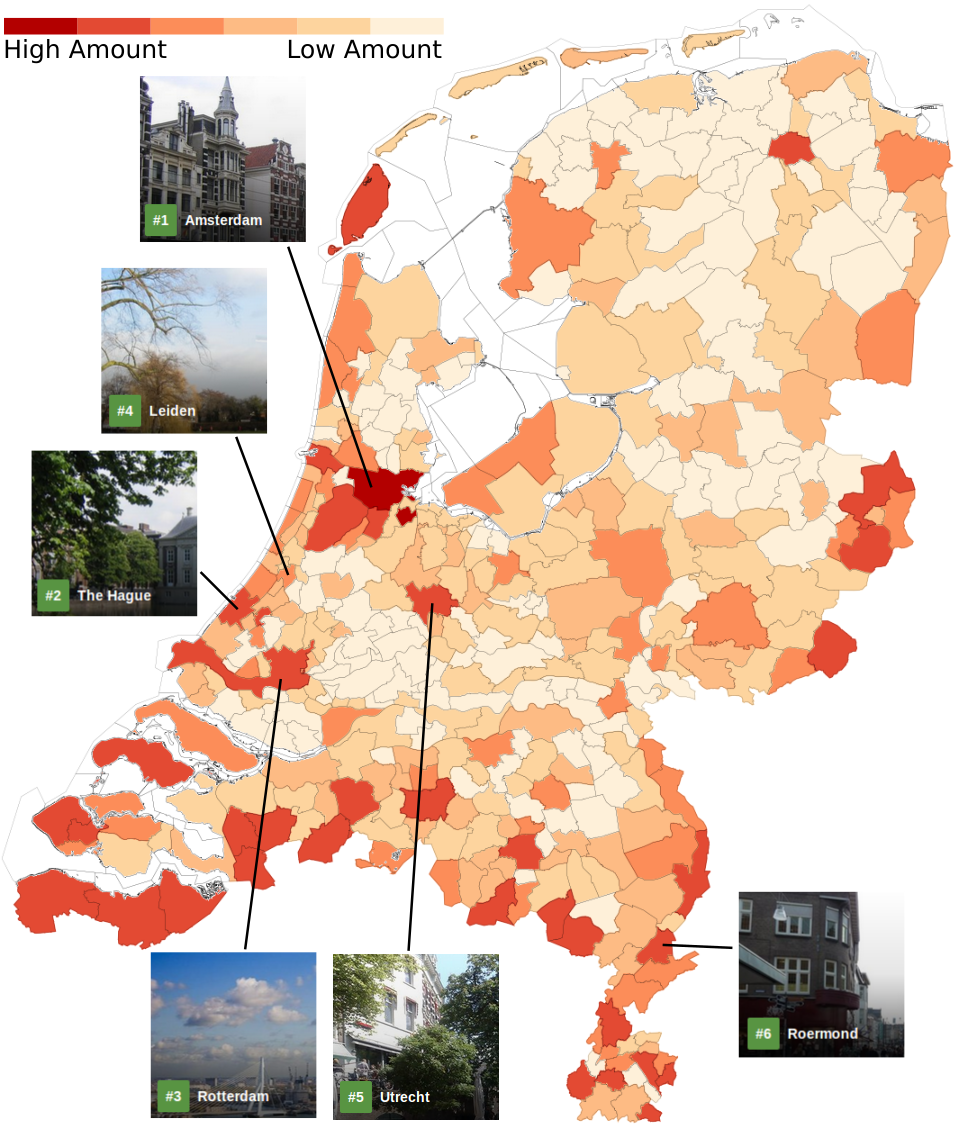}
}{
  \caption{The most popular destinations in Netherlands (red = highest USD expenditure, white = low/no USD expenditure). Data corresponds to Table \ref{tab:nld-destinations}, color map is in logs. We highlight the top 6 tourism destinations according to Trip Advisor. Mastercard insights.}
  \label{fig:nld-destinations}
}
\capbtabbox{
\begin{tabular}{l|l|rr}
Rank & Zone & \%\\
\hline
1 & Amsterdam & 18.84\%\\
2 & Roermond & 6.61\%\\
3 & Maastricht & 5.67\%\\
4 & Haarlemmermeer & 4.74\%\\
5 & Venlo & 3.44\%\\
6 & Amstelveen & 2.60\%\\
7 & Heerlen & 2.53\%\\
8 & Sluis & 2.47\%\\
9 & Breda & 2.46\%\\
10 & Rotterdam & 2.15\%\\
11 & 'S-Gravenhage & 2.10\%\\
12 & Hulst & 2.03\%\\
13 & Eindhoven & 1.79\%\\
14 & Enschede & 1.04\%\\
15 & Veere & 1.03\%\\
16 & Roosendaal & 1.01\%\\
17 & Terneuzen & 1.00\%\\
18 & Woensdrecht & 0.96\%\\
19 & Weert & 0.93\%\\
20 & Winterswijk & 0.89\%\\
\end{tabular}
}{
\caption{The top tourist destinations for Netherlands in percentage of dollars spent inside the country across all countries of origin, industries and time periods. Mastercard insights.}
\label{tab:nld-destinations}
}
\end{floatrow}
\end{figure}

Figure \ref{fig:nld-destinations} and Table \ref{tab:nld-destinations} show the geographical distribution of tourist expenditures in the Netherlands. We can make mostly the same observations as for Colombia about the three classes of the main attractors. However, there is a major difference between the two countries. The Netherlands has a higher density of both population (493 per km$^2$ versus Colombia's 42), places of interest and large densely populated nearby neighbors. This fact is reflected in the distribution of expenditures in the country: virtually every municipality in the Netherlands has been visited and has seen an influx of foreign payment card expenditure.

\begin{figure}[!h]
\centering
\includegraphics[width=.495\columnwidth]{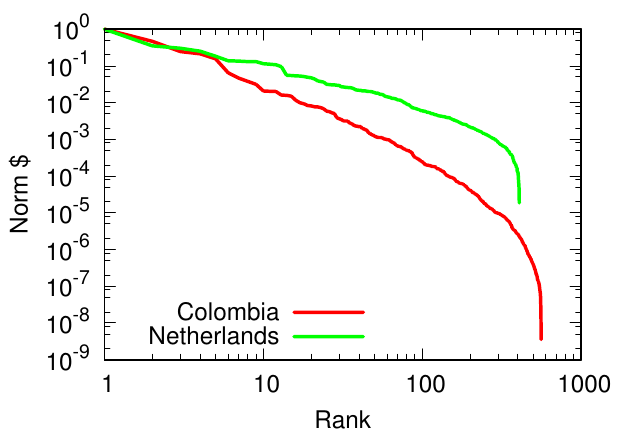}
\caption{The distribution of foreign card expenditures per municipality of destination, for Colombia and the Netherlands. Cities are ranked on the x axis in descending order of expenditure normalized by the maximum expenditure (y axis). Mastercard insights.}
\label{fig:dest-distr}
\end{figure}

We show these distributions in Figure \ref{fig:dest-distr}. The distribution for the Netherlands is flatter. The 75th percentile municipality earned only 7 times as much as the 25th percentile municipality. This ratio for Colombia is 57. This evidence suggests that tourism and foreign expenditure are driven in two radically different ways in the countries we are studying. Colombia has few centralized attractors, while the Netherlands is more akin to a decentralized self-organizing system. 

\subsection{Where are the tourists coming from?}
\begin{figure}[!t]
\begin{floatrow}
\capbtabbox{
\begin{tabular}{l|l|r}
Rank & Country & \%\\
\hline
1 & United States & 33.92\%\\
2 & Venezuela & 11.00\%\\
3 & Netherlands & 3.97\%\\
4 & Germany & 3.96\%\\
5 & Switzerland & 3.93\%\\
6 & Chile & 3.82\%\\
7 & Mexico & 3.66\%\\
8 & Australia & 3.06\%\\
9 & Spain & 2.97\%\\
10 & Italy & 2.90\%\\
\end{tabular}
}
{
\caption{The top ten tourist origins for Colombia in percentage of dollars spent inside the country across all municipalities, industries and time periods. Mastercard insights.}
\label{tab:col-origins}
}
\capbtabbox{
\begin{tabular}{l|l|r}
Rank & Country & \%\\
\hline
1 & Belgium & 30.23\%\\
2 & Germany & 25.64\%\\
3 & United Kingdom & 4.68\%\\
4 & United States & 4.39\%\\
5 & Poland & 3.99\%\\
6 & Switzerland & 3.72\%\\
7 & France & 2.42\%\\
8 & Italy & 2.26\%\\
9 & Russia & 1.91\%\\
10 & Australia & 1.72\%\\
\end{tabular}
}
{
\caption{The top ten tourist origins for Netherlands in percentage of dollars spent inside the country across all municipalities, industries and time periods. Mastercard insights.}
\label{tab:nld-origins}
}
\end{floatrow}
\end{figure}

\begin{figure}
\centering
\includegraphics[width=\columnwidth]{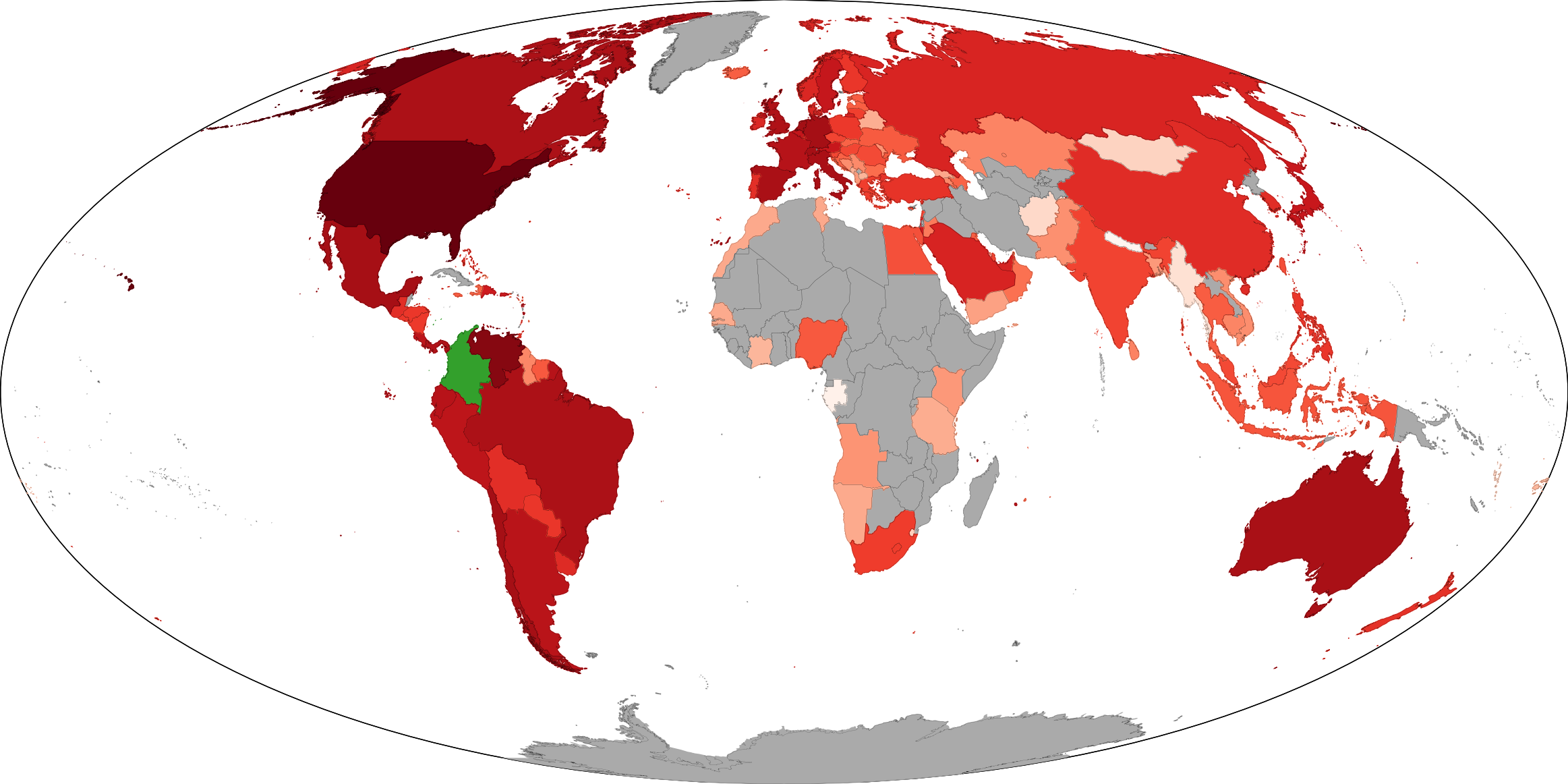}\\
\vspace{1cm}
\includegraphics[width=\columnwidth]{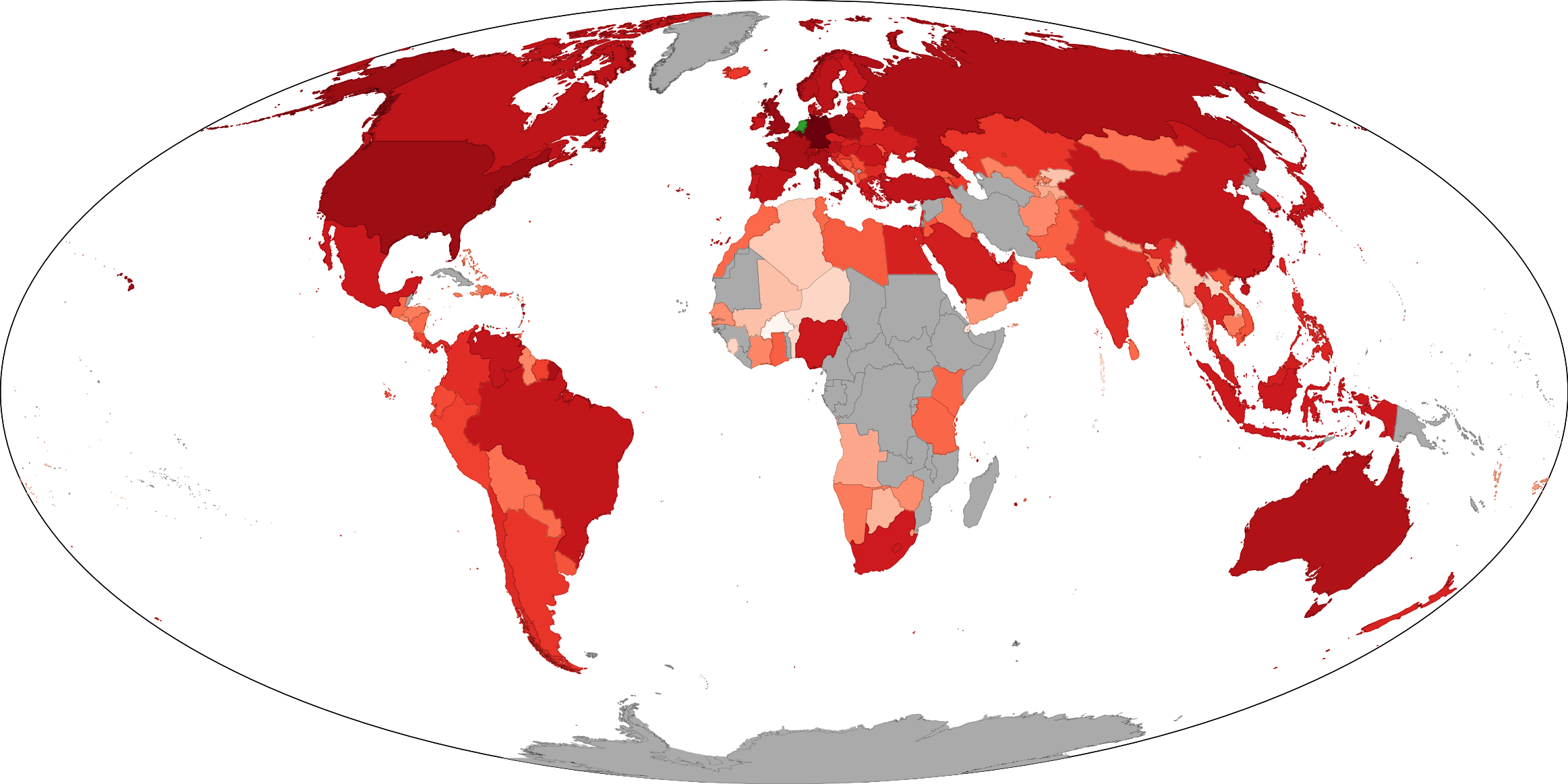}
\caption{The global map of foreign card origins traveling to Colombia (top) and to the Netherlands (bottom). The color map is in logs: red = highest USD expenditure, white = low USD expenditure, gray = no data -- country not included in the sample --, green = country of destination. Mastercard insights.}
\label{fig:orig-map}
\end{figure}

Tables \ref{tab:col-origins} and \ref{tab:nld-origins} and Figure \ref{fig:orig-map} report on the origins of the foreign cards making expenditures in Colombia and the Netherlands. In the heat map we illustrate the countries' consumption to foreign spend in Columbia and the Netherlands. In the tables we report the share of expenditure corrected for the total population of the country of origin. The influence of geographical distance is evident, and it highlights again the importance of size of the country and the population density of the countries nearby. The second partner for Colombia is Venezuela, the first and second partners for the Netherlands are Belgium and Germany: all bordering countries. However, longer range trips still may play an important role: we note significant expenditures in the Netherlands by US tourists, and the Netherlands itself is the third overall tourism origin for Colombia.

We can test the importance of geographical distance, along with other distance measures, in the destination decision of tourists. This can be done creating a simple gravity model as follows:

$$ \log(E_{o,d}) = \alpha + \rho D_{o,d} + \beta X_o + \epsilon_{o,d}, $$

where $o$ and $d$ are the country of origin and country of destination respectively, $E_{o,d}$ is the amount of dollars spent in $d$ by cards issued in $o$, $X_o$ is a set of variables measuring the size of the origin, $D_{o,d}$ a set of reciprocal distance variables between $o$ and $d$, $\alpha$ is a constant and $\epsilon$ the error term.

\begin{table}[!h] \centering
\begin{tabular}{@{\extracolsep{5pt}}lD{.}{.}{-3} D{.}{.}{-3} | D{.}{.}{-3} D{.}{.}{-3} } 
\\[-1.8ex]\hline 
\hline \\[-1.8ex] 
 & \multicolumn{4}{c}{\textit{Dependent variable:}} \\ 
\cline{2-5}
\\[-1.8ex] & \multicolumn{4}{c}{$\log(E_{o,d})$} \\ 
\\[-1.8ex] & \multicolumn{2}{c|}{Colombia} & \multicolumn{2}{c}{Netherlands} \\ 
\\[-1.8ex] & \multicolumn{1}{c}{(1)} & \multicolumn{1}{c|}{(2)} & \multicolumn{1}{c}{(3)} & \multicolumn{1}{c}{(4)}\\ 
\hline \\[-1.8ex] 
 $\log(POP_o)$ & 1.110^{***} & 1.037^{***} & 0.989^{***} & 0.857^{***} \\ 
  & (0.085) & (0.093) & (0.089) & (0.097) \\ 
  & & & & \\ 
 $\log(GDPPC_o)$ & 2.017^{***} & 2.008^{***} & 1.915^{***} & 1.727^{***} \\ 
  & (0.107) & (0.110) & (0.114) & (0.122) \\ 
  & & & & \\ 
 $\log(D_{o,d})$ & -2.292^{***} & -1.712^{***} & -0.544^{***} & -0.244 \\ 
  & (0.180) & (0.266) & (0.169) & (0.178) \\ 
  & & & & \\ 
 Common Language &  & 1.647^{***} &  & 2.605^{**} \\ 
  &  & (0.573) &  & (1.168) \\ 
  & & & & \\ 
 $\log(F_{o,d})$ &  & -0.002 &  & 0.107^{***} \\ 
  &  & (0.044) &  & (0.033) \\ 
  & & & & \\ 
 Constant & -2.744 & -6.955^{***} & -13.844^{***} & -13.249^{***} \\ 
  & (2.231) & (2.573) & (2.563) & (2.490) \\ 
  & & & & \\ 
\hline \\[-1.8ex] 
Observations & \multicolumn{1}{c}{117} & \multicolumn{1}{c|}{117} & \multicolumn{1}{c}{135} & \multicolumn{1}{c}{135} \\ 
R$^{2}$ & \multicolumn{1}{c}{0.822} & \multicolumn{1}{c|}{0.836} & \multicolumn{1}{c}{0.781} & \multicolumn{1}{c}{0.807} \\ 
Adjusted R$^{2}$ & \multicolumn{1}{c}{0.817} & \multicolumn{1}{c|}{0.828} & \multicolumn{1}{c}{0.776} & \multicolumn{1}{c}{0.799} \\ 
Residual Std. Error & \multicolumn{1}{c}{1.479} & \multicolumn{1}{c|}{1.433} & \multicolumn{1}{c}{1.691} & \multicolumn{1}{c}{1.602} \\ 
F Statistic & \multicolumn{1}{c}{173.513$^{***}$} & \multicolumn{1}{c|}{112.793$^{***}$} & \multicolumn{1}{c}{155.777$^{***}$} & \multicolumn{1}{c}{107.656$^{***}$} \\ 
\hline 
\hline \\[-1.8ex] 
\textit{Note:}  & \multicolumn{4}{r}{$^{*}$p$<$0.1; $^{**}$p$<$0.05; $^{***}$p$<$0.01} \\ 
\end{tabular}
\caption{The result of the gravity models for Colombia and the Netherlands. Mastercard insights, integrated with publicly available data about population and GDP of countries, and CEPII's country-country distance measurements.} 
\label{tab:gravity-model} 
\end{table}

Table \ref{tab:gravity-model} reports the results of such models. We run the model for Colombia and for the Netherlands separately, so there is no need to control for the size of the destination. We run two models for each country, in which we disaggregate the size and the distance variables in different ways. Models 1 and 2 refer to Colombia, and models 3 and 4 to the Netherlands. Models 1 and 3 use two variables for $X_o$: population and GDP per capita of the country of origin. They use only the geographical distance as $D_{o,d}$. It is calculated as a weighted combined distance between all their major cities -- as supplied in \cite{cepiigeo}. In models 2 and 4 we introduce two corrections for the distance variables: whether the two countries share a language and how strong flight connections between them are. These variables should provide a control for cultural affinity and actual travel effort.

For both Colombia and the Netherlands the size of the origin matters in comparable amounts. Countries with higher GDP per capita and population have greater expenditures in their destinations. Distance has the expected negative sign: countries farther away are less likely to visit the destinations -- thus lowering the total amount of expenditures. The effect seems to be particularly strong for Colombia, for which the coefficient is more than four times as high as in the Netherlands. Note that these models already achieve an $R^2$ of around 80\%, showing that these three variables allow for useful insights as to where tourism flows, providing an important validation about the robustness of the anonymized and aggregated transaction data.

When adding corrections for the distance variable in models 2 and 4, we note that the two countries experience very different dynamics. While the language variable appears to have a stronger effect in the Netherlands, its significance is lower. This effect might be due to the different popularity of the two languages across the world -- there are around half a billion Spanish native speakers in the world, while Dutch has fewer than 30 millions. On the other hand, direct flight connections seem to have no effect for Colombia, while in the Netherlands they nullify the effect of geographical distance. this indicates the importance of being a hub in the world air transport network.

\begin{figure}
\centering
\includegraphics[width=.495\columnwidth]{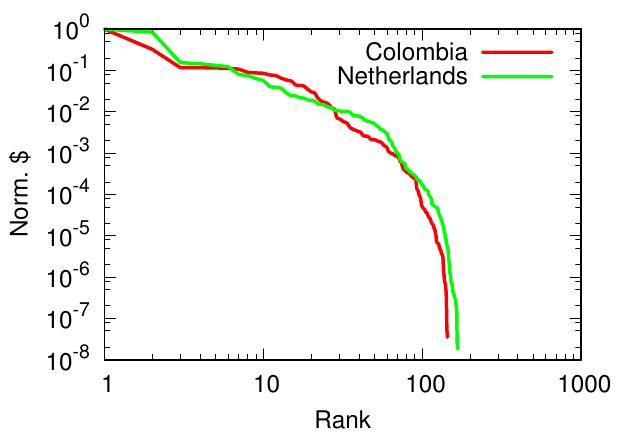}
\caption{The distribution of foreign card expenditures per country of origin, for Colombia and the Netherlands. Countries are ranked on the x axis in descending order of expenditure normalized by the maximum expenditure (y axis). Mastercard insights.}
\label{fig:orig-distr}
\end{figure}

\begin{figure}
\begin{floatrow}
\capbtabbox{
\begin{tabular}{l|l|r}
Rank & Industry & \%\\
\hline
1 & ATMs & 54.38\%\\
\hline
2 & Accommodations & 13.64\%\\
3 & Grocery Stores & 11.75\%\\
4 & Eating Places & 7.39\%\\
5 & Family Apparel & 7.01\%\\
6 & T+E Airlines & 6.48\%\\
7 & Public Administration & 6.21\%\\
8 & Miscellaneous & 5.58\%\\
9 & Drug Store Chains & 5.35\%\\
10 & Jewelry and Giftware & 3.92\%\\
\end{tabular}
}
{
\caption{The top ten industries for Colombia in percentage of expenditures across all origins, municipalities and time periods. For ATM we report the percentage of the total expenditures, for other industries the percentage of the non-ATM expenditures. Mastercard insights.}
\label{tab:col-industries}
}
\capbtabbox{
\begin{tabular}{l|l|r}
Rank & Industry & \%\\
\hline
1 & ATMs & 28.08\%\\
\hline
2 & Accommodations & 12.25\%\\
3 & Grocery Stores & 11.11\%\\
4 & Family Apparel & 8.92\%\\
5 & Eating Places & 6.78\%\\
6 & Automotive Fuel & 6.51\%\\
7 & Wholesale Trade & 4.29\%\\
8 & Home Furnishings / Furniture & 3.93\%\\
9 & T+E Airlines & 3.86\%\\
10 & Sporting Goods / Apparel / Footwear & 2.61\%\\
\end{tabular}
}
{
\caption{The top ten industries for Netherlands in percentage of expenditures across all origins, municipalities and time periods. For ATM we report the percentage of the total expenditures, for other industries the percentage of the non-ATM expenditures. Mastercard insights.}
\label{tab:nld-industries}
}
\end{floatrow}
\end{figure}

\begin{figure}
\centering
\includegraphics[width=.29\columnwidth]{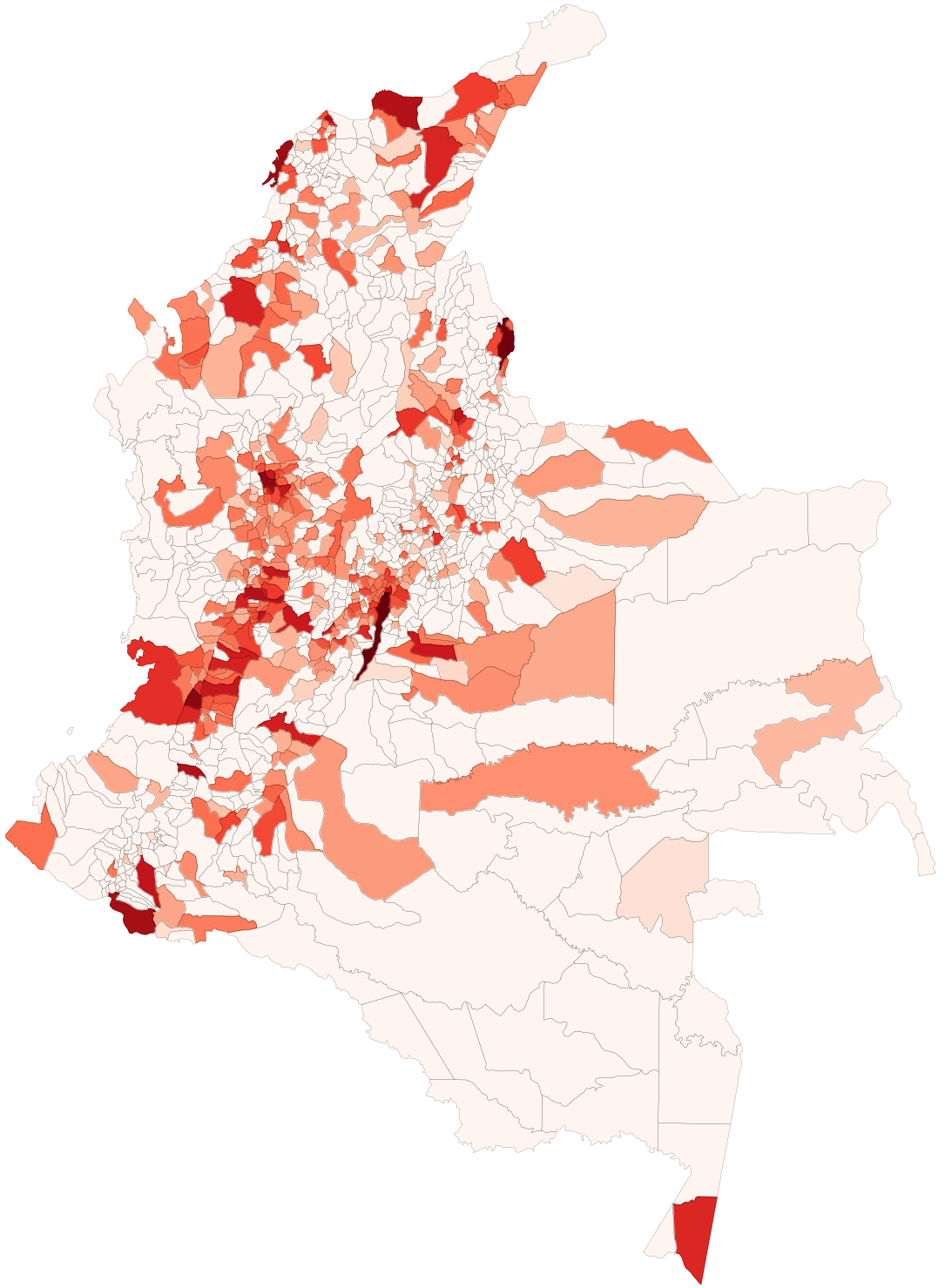}
\includegraphics[width=.29\columnwidth]{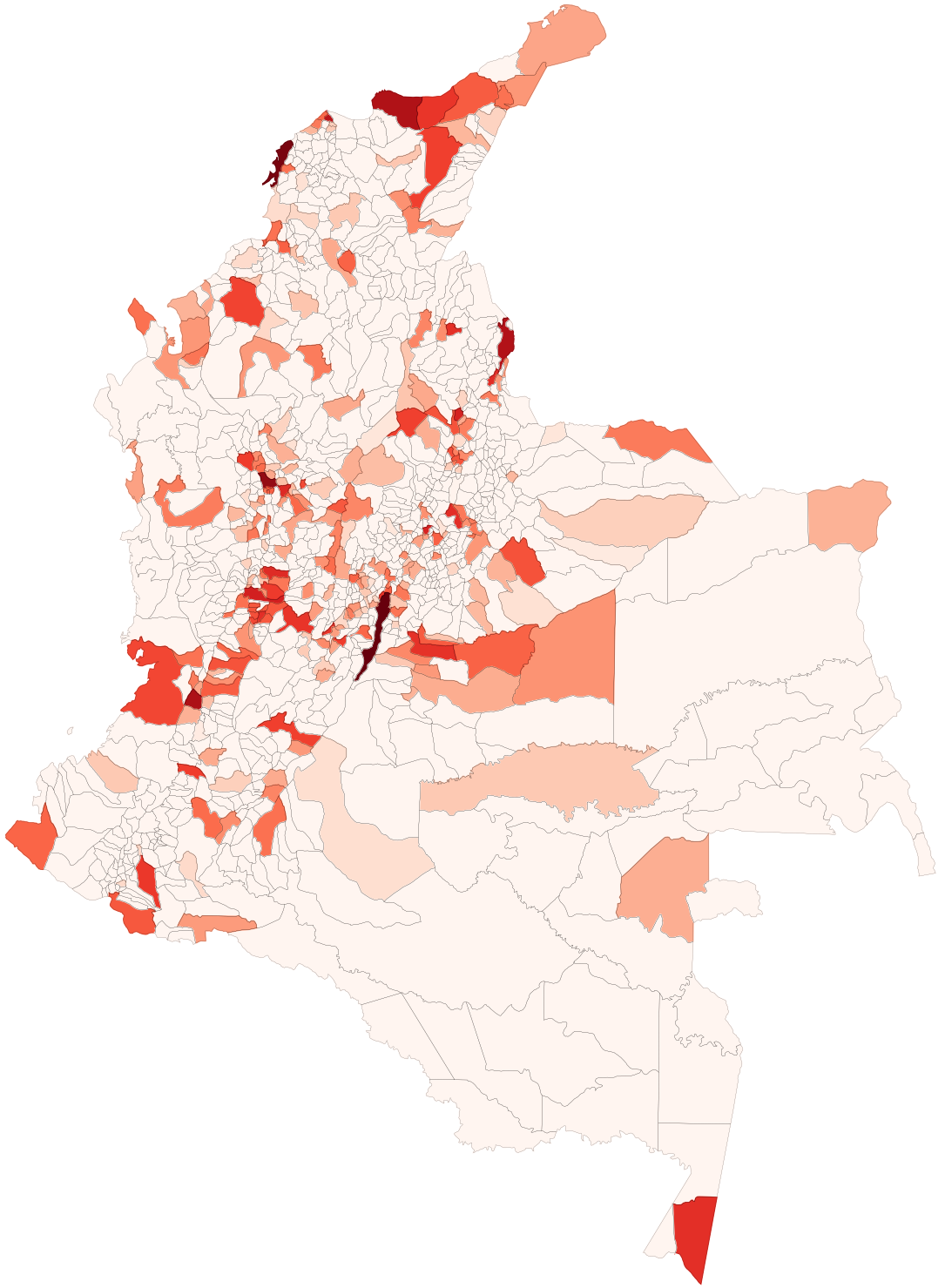}
\includegraphics[width=.29\columnwidth]{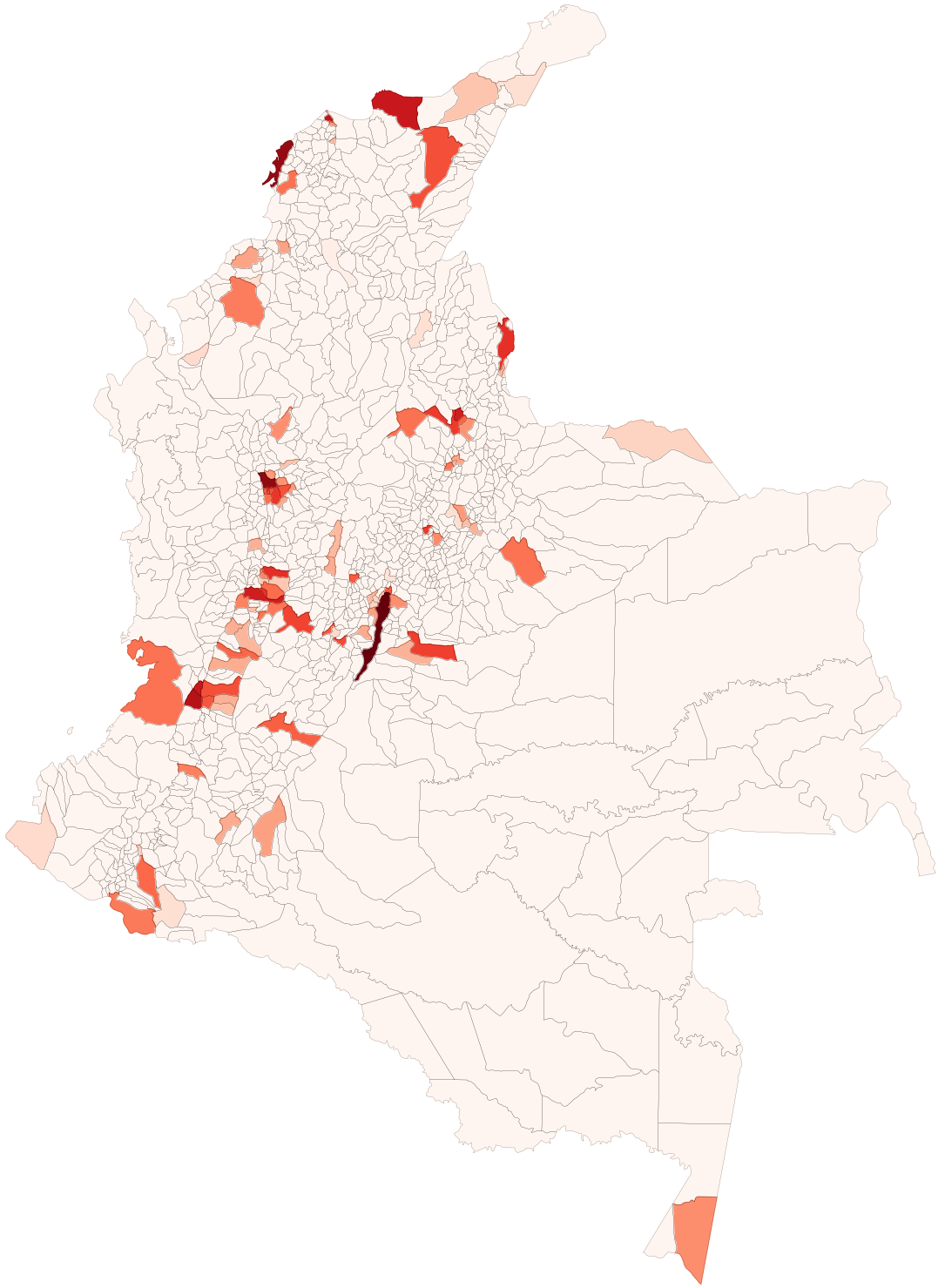}
\includegraphics[width=.29\columnwidth]{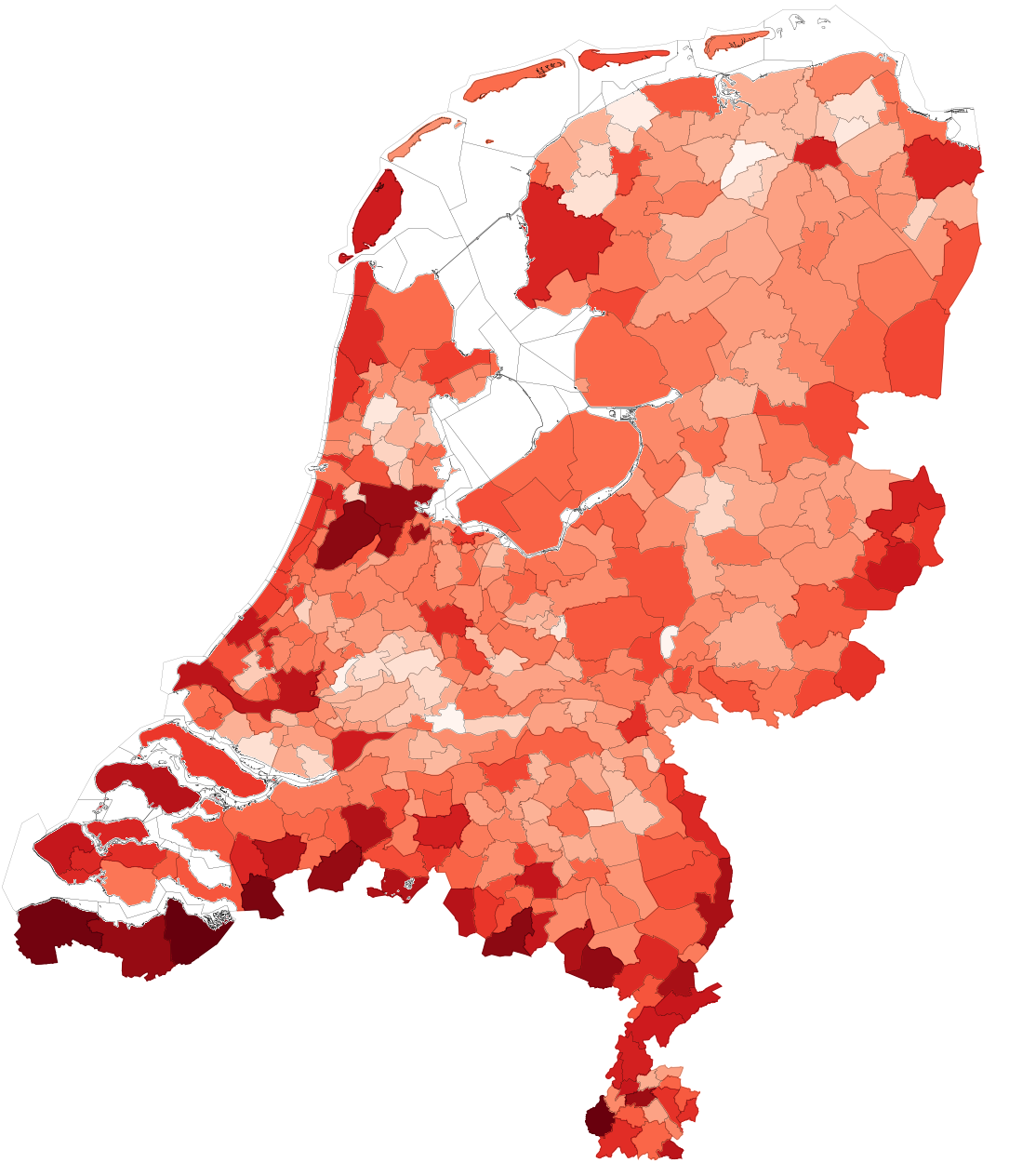}
\includegraphics[width=.29\columnwidth]{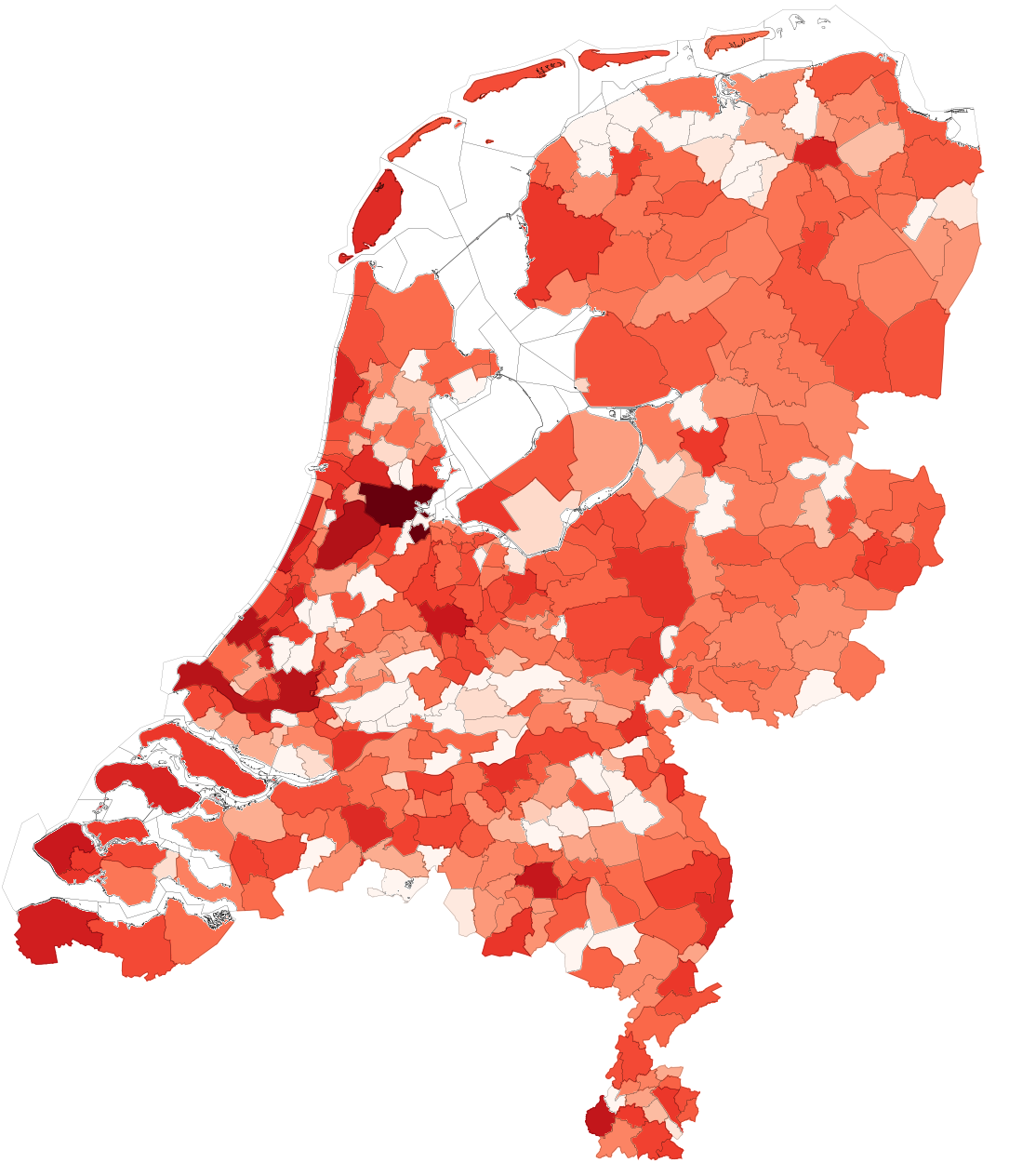}
\includegraphics[width=.29\columnwidth]{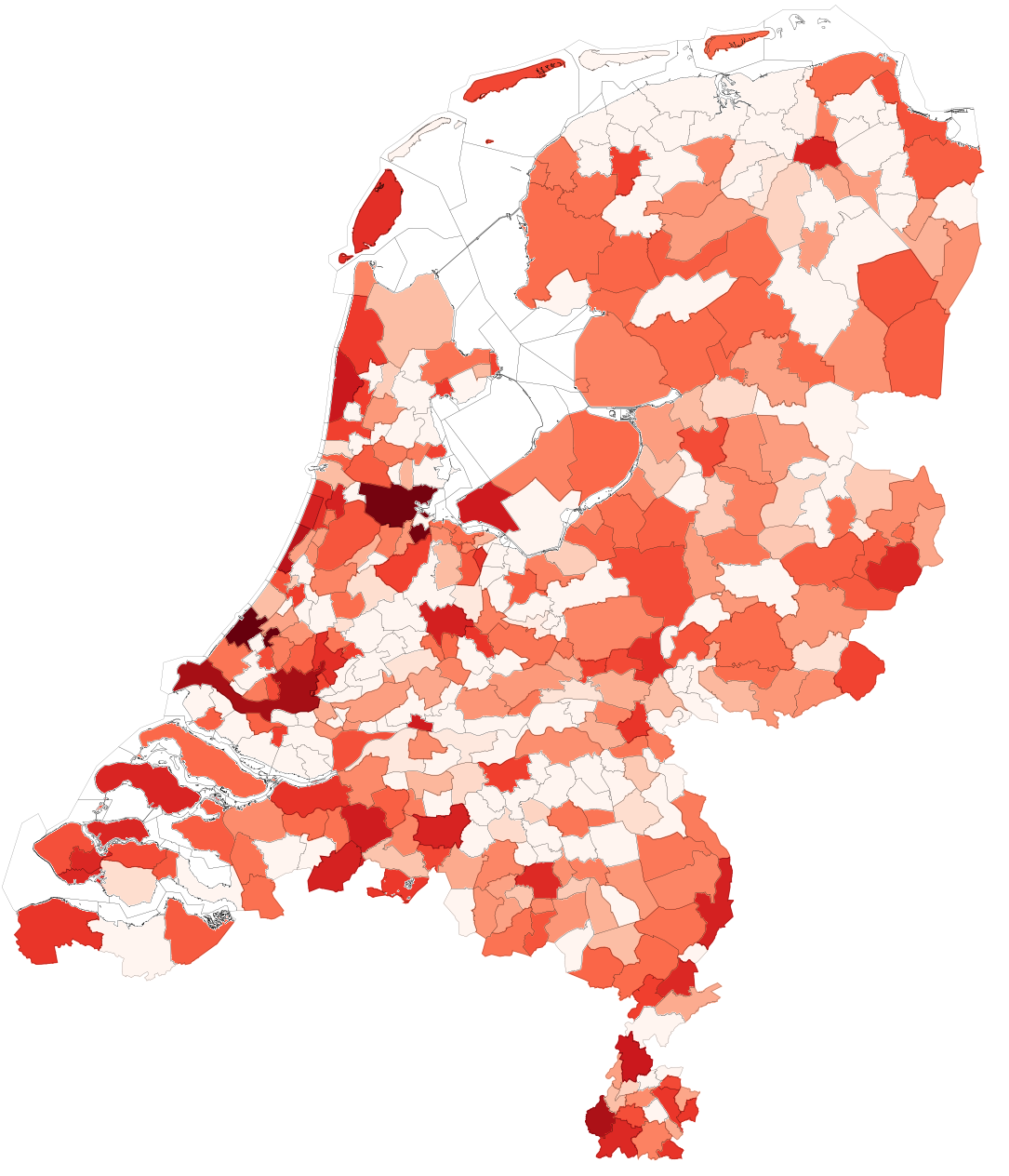}
\caption{Sector-specific maps for Colombia (top) and the Netherlands (bottom). From left to right: grocery stores, accommodations and bars/taverns/nightclubs. Mastercard insights.}
\label{fig:sectormaps}
\end{figure}

When looking at the distribution of expenditures per country (Figure \ref{fig:orig-distr}) we do not see the difference in slope we observed for the municipality of destination distribution (Figure \ref{fig:dest-distr}) -- only a difference in scale. This means that both Colombia and the Netherlands experience similar relationships with their tourist origins. The degree with which they rely on few large originators or many small ones is about the same.

\subsection{What do tourists buy?}
The data allow us to look at foreign card spend by country. Each merchant is required to classify its activities using a three digit system. Tables \ref{tab:col-industries} and \ref{tab:nld-industries} report the top ten merchant types by expenditure in Colombia and the Netherlands. The first takeaway is that cash usage is more prominent in Colombia: more than half the foreign currency exchanges in the country happened through ATMs. The share for the Netherlands is less than a third.

The top four merchant types besides ATMs are the same in the two countries, and correspond to common activities associated with tourism: hotels (accommodations), dining (eating places) and shopping (family apparel). It is important to note that the activity of buying products at grocery stores is not traditionally associated with tourism, but here it ranks as third highest in terms of volume. In the Netherlands it touches a billion dollars. This activity is probably in part due to commuters rather than tourists. However it is still an export of the country, and one that is likely not to be captured by traditional tourism estimation methodologies.

That tourists visit grocery stores is confirmed when comparing sector-specific expenditure maps. Figure \ref{fig:sectormaps} reports six of them. Grocery stores expenditures are spread all over the Colombian/Dutch territory, even very far from the borders. By comparison, the expenditures for accommodation are more concentrated in large cities. Sector-specific maps are also useful to characterize other activities as spatially concentrated or not. In the case of the Netherlands, a tourist can find a bar in every large municipality of the country. By contrast, in Colombia bars visited by tourists are more spatially concentrated.

\subsection{When do tourists come?}
\begin{figure}
\centering
\includegraphics[width=.495\columnwidth]{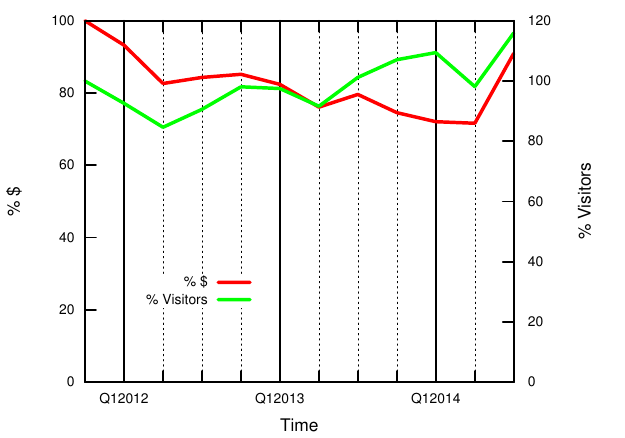}
\includegraphics[width=.495\columnwidth]{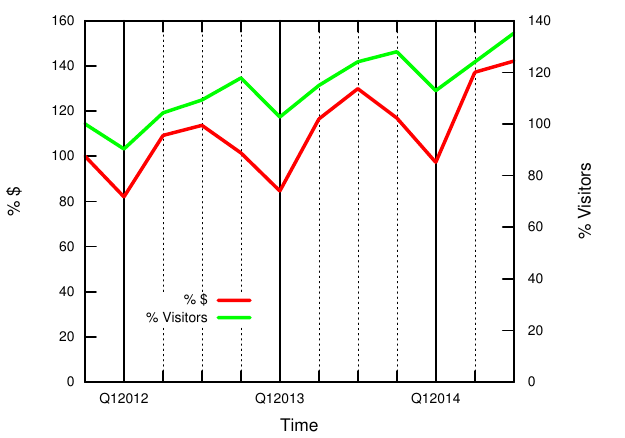}
\caption{The evolution of foreign card expenditures and visits for Colombia (left) and Netherlands (right). For each quarter, the reported sums are totals, aggregated across all origins, destinations and industries. We did not perform any seasonal adjustment. Mastercard insights.}
\label{fig:timelines}
\end{figure}

\begin{figure}[!t]
\centering
\includegraphics[width=.495\columnwidth]{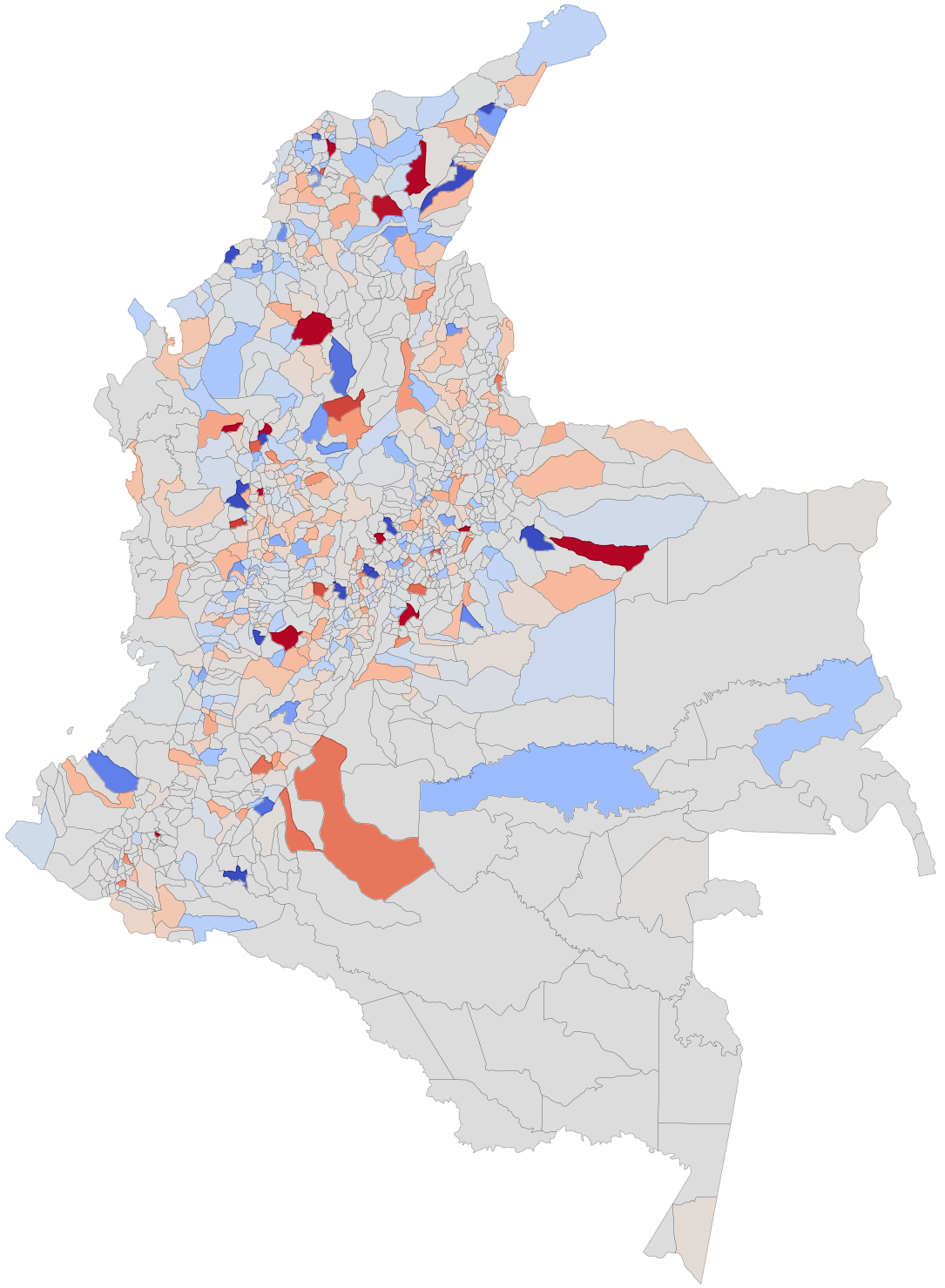}
\includegraphics[width=.495\columnwidth]{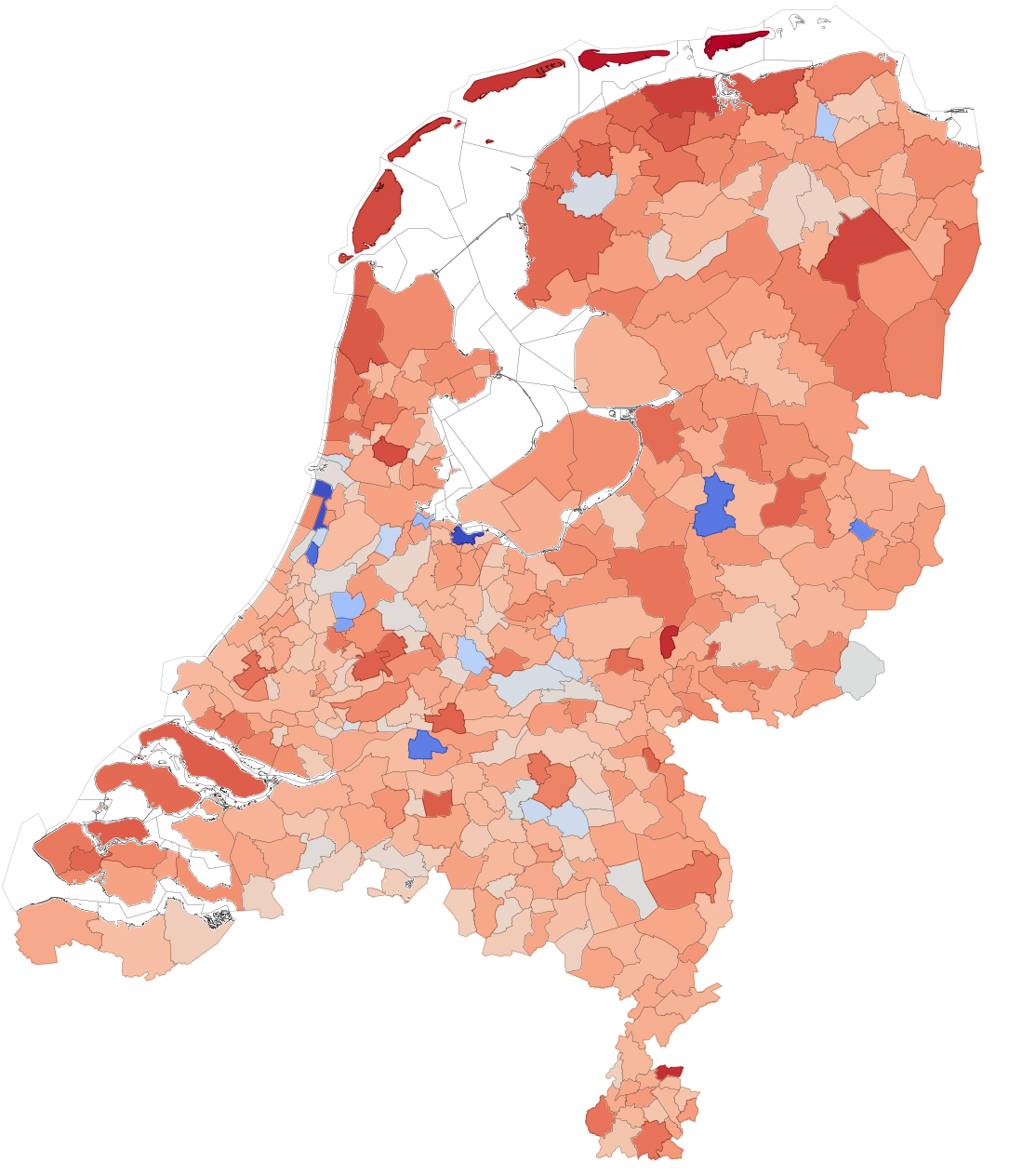}
\caption{The quarter expenditure balance in Colombia (left) and the Netherlands (right). Each municipality is colored according to its tendency of attracting expenditures in summer (red) and in winter (blue). If there is no difference between the two season the municipality is colored in gray. Mastercard insights.}
\label{fig:time-maps}
\end{figure}

Figure \ref{fig:timelines} reports, for each quarter, the total foreign card expenditure and visits. There are many lessons learned from these plots. The first is that the two countries seem to be on different trends. With the exception of the last spike, Colombia is declining in total foreign activity. The number of foreign visitors is increasing, but the overall revenue is decreasing. The Dutch trend, besides having overall larger levels under any of the criteria discussed so far, is also upward. This diagnostic test suggests that Colombia has to fix a struggling sector.

The second lesson learned focuses on the two different quarterly patterns. Colombia has smoother transitions from quarter to quarter. The seasonal effect is instead very prominent for the Netherlands: the first quarter of the year, winter, is always the least popular. Tourism expenditures peak in summer, the third quarter. This might have to do with the different position of the two countries: being placed near the equator Colombia has an isothermal climate, so it can attract tourists equally across all quarters. Winters in the Netherlands are not harsh, but with an expected difference in average temperature of $15^{\circ}C$ between January and July -- and no mountains to attract tourists engaged in winter activities -- a significant drop is to be expected.

Again, this is not a discovery, rather a validation of seasonal effects. We also visualize this on a map. Figure \ref{fig:time-maps} reports the seasonal balance for each municipality in the two countries. We can confirm that in Colombia most of the popular municipalities show no difference between summer and winter. Most changes happen in municipalities that are not very popular, i.e. outliers. The situation is radically different in the Netherlands, as expected. Most municipalities are popular in summer, and winter shows little to no preference across the country. This analysis is useful to classify the tourist activity as being seasonal or year-round also at a subnational level, if necessary.

\section{Enhancing Tourism}
After the previous descriptive section, we now turn our attention to a possible prescriptive application. In this section we outline a possible basis for an analytic framework for enhancing the tourism possibilities of countries and municipalities. We look at three slices of the anonymized and aggregated transaction data, by estimating a correlation-based similarities of origins, destinations and industries. These are just the starting points of what can be easily implemented as a tourism enhancer collaborative filtering system, akin what Amazon and Netflix already do to promote their products \cite{linden2003amazon, koren2010collaborative}.

Note that for this section we introduce foreign credit/debit expenditures for four additional countries: Albania, Greece, Slovenia and Croatia. We do so to increase the robustness of our insights.

\subsection{Origin Space}
\begin{figure}[!b]
\centering
\includegraphics[width=.8\columnwidth]{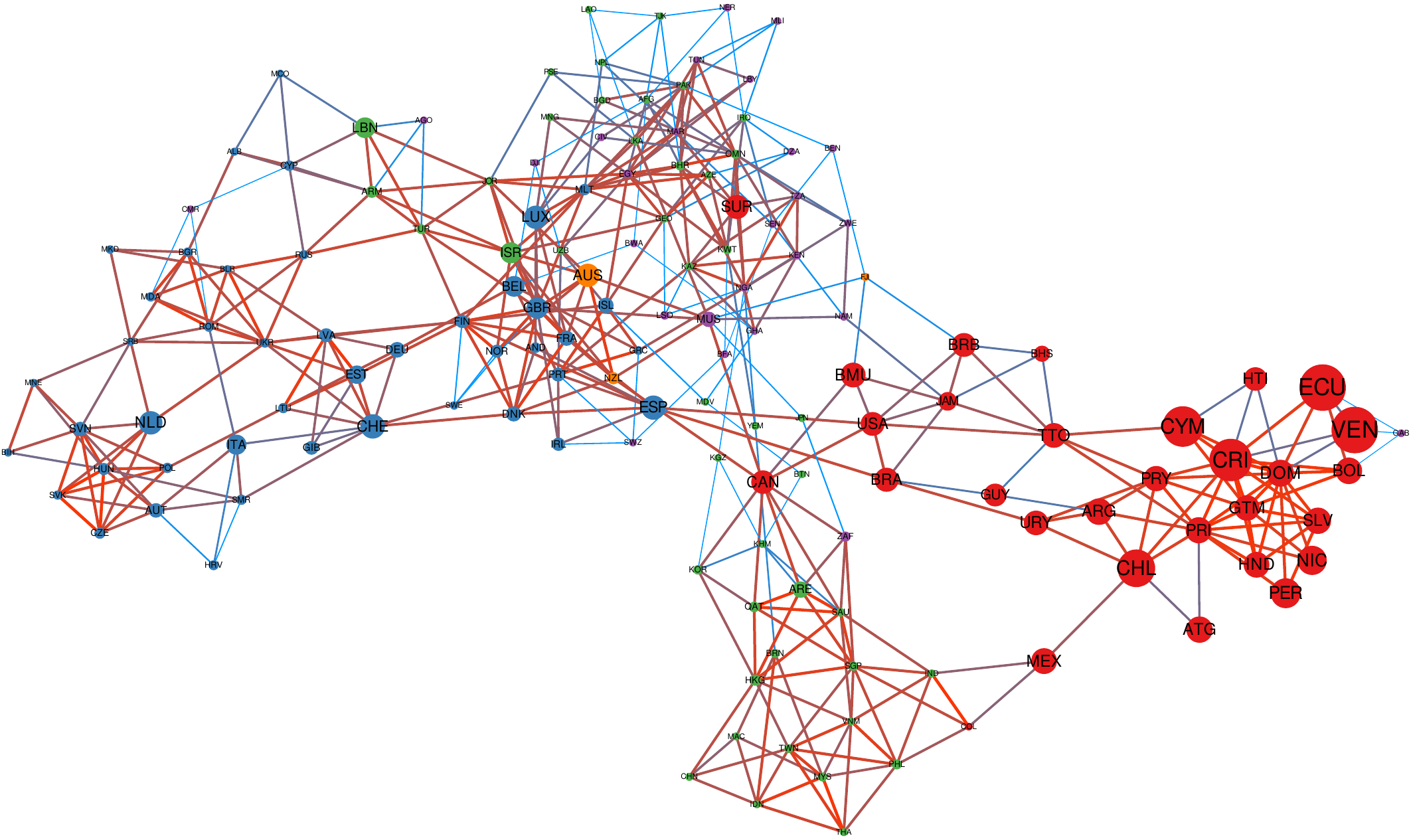}
\includegraphics[width=.8\columnwidth]{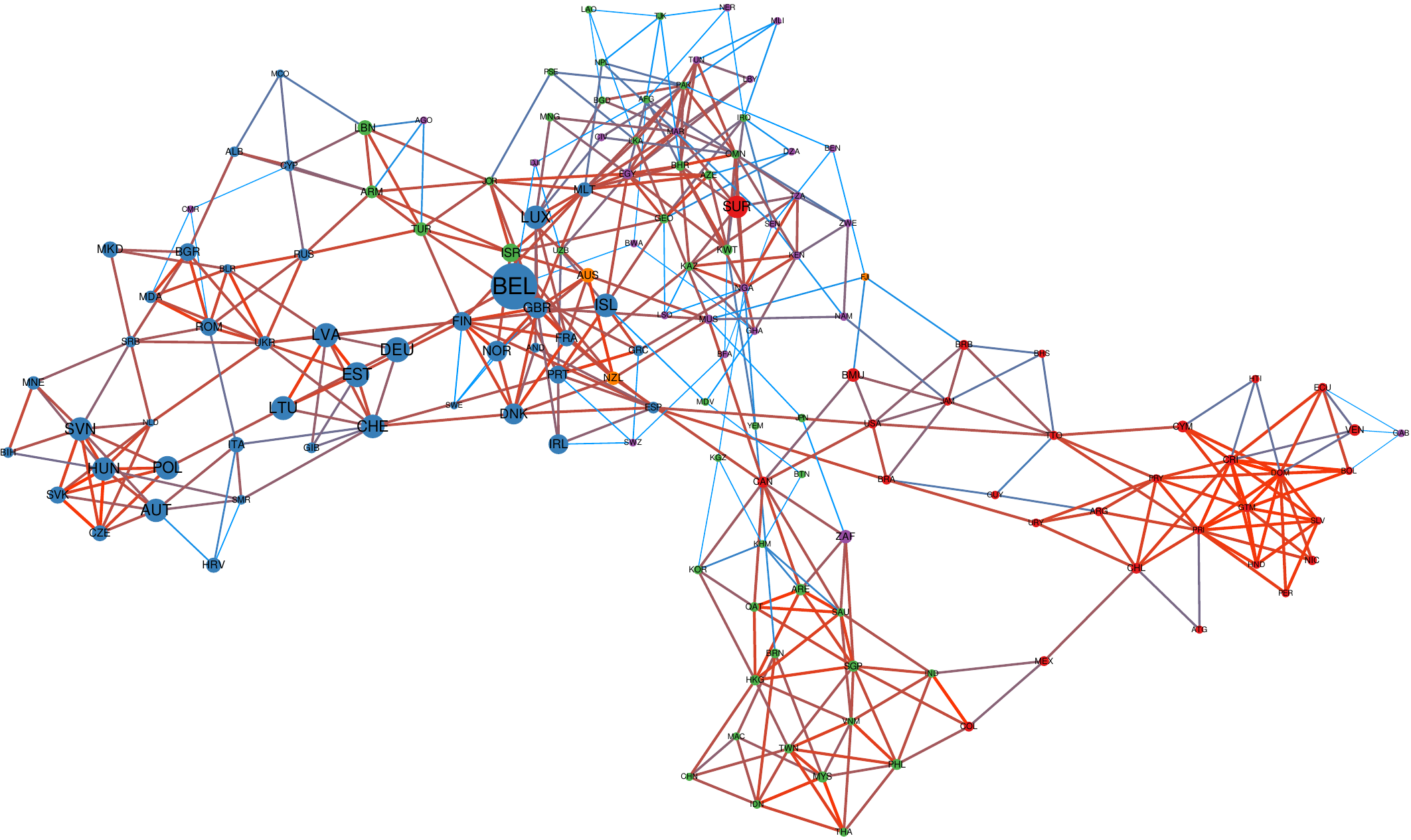}
\caption{The Origin Space for Colombia (top) and the Netherlands (bottom). Each node is a country. Countries are connected if there is a high correlation in the destination where their tourists spend their money. The country's continent determines the node color. Edge size and color is proportional to the country-country similarity. The topology of the edges is the same for Colombia and the Netherlands. The node size is proportional to the relative attractivness of the destination for the origin. It is determined by dividing the number of dollars spent in the destination with the origin's GDP, normalized by the total shares in the destination. Mastercard insights, integrated by public data about the GDP of countries.}
\label{fig:country-spaces}
\end{figure}

We start by building what we call the ``Origin Space'', a network connecting two countries if their tourists have similar expenditure patterns. For each origin, we build a vector containing the total dollar expenditure of cards issued in that origin in each of the six possible countries of destination. Each entry in the vector is aggregated across all industries $I$, all municipalities $M_d$ and all quarters $Q$:

$$ V_o = \left \{ \sum_{i, m, q} E_{o,i,m,q}, \forall d \in D \right \},$$

where $o$ is the country of origin, $d$ is the country of destination, $M_d$ is the set of municipalities in $d$ and $E_{o,i,m,q}$ is the amount spent in dollars in industry $i \in I$, municipality $m \in M_d$ and quarter $q \in Q$.

We visualize the Origin Space as a network. In Figure \ref{fig:country-spaces}, we show only three connections per country, connecting it with the three countries to which it is the most similar. Countries can have more than three connections because more countries are similar to them. The Origin Space has a strong geographical component: origins that are close to each other are likely to have similar tourism patterns. This is a direct consequence of the distance effect in travel, as noted in Table \ref{tab:gravity-model}.

\begin{table}[!b] \centering \small 
\begin{tabular}{@{\extracolsep{5pt}}lD{.}{.}{-3} } 
\\[-1.8ex]\hline 
\hline \\[-1.8ex] 
 & \multicolumn{1}{c}{\textit{Dependent variable:}} \\ 
\cline{2-2} 
\\[-1.8ex] & \multicolumn{1}{c}{$\ln(E_{o,d} + 1)$} \\ 
\hline \\[-1.8ex] 
 $\ln(P_{o,d} + 1)$ & 1.850^{***} \\ 
  & (0.122) \\ 
  & \\ 
 Constant & -16.411^{***} \\ 
  & (1.667) \\ 
  & \\ 
\hline \\[-1.8ex] 
Origin FE & \multicolumn{1}{c}{Y} \\ 
Destination FE & \multicolumn{1}{c}{Y} \\ 
\hline \\[-1.8ex] 
Observations & \multicolumn{1}{c}{898} \\ 
R$^{2}$ & \multicolumn{1}{c}{0.874} \\ 
Adjusted R$^{2}$ & \multicolumn{1}{c}{0.848} \\ 
Residual Std. Error & \multicolumn{1}{c}{2.118} \\ 
F Statistic & \multicolumn{1}{c}{33.187$^{***}$} \\ 
\hline 
\hline \\[-1.8ex] 
\textit{Note:}  & \multicolumn{1}{r}{$^{*}$p$<$0.1; $^{**}$p$<$0.05; $^{***}$p$<$0.01} \\ 
\end{tabular} 
\caption{The relationship between the observed foreign card expenditures in a destination and what we would expect given the origin-origin correlations in the Origin Space. Mastercard insights.} 
\label{tab:country-space-prediction-1} 
\end{table}

\begin{table}[!!!h] \centering \small 
\begin{tabular}{@{\extracolsep{5pt}}lD{.}{.}{-3} D{.}{.}{-3} D{.}{.}{-3} } 
\\[-1.8ex]\hline 
\hline \\[-1.8ex] 
 & \multicolumn{3}{c}{\textit{Dependent variable:}} \\ 
\cline{2-4} 
\\[-1.8ex] & \multicolumn{3}{c}{$\ln(E^{t+1}_{o,d})$} \\ 
\\[-1.8ex] & \multicolumn{1}{c}{(1)} & \multicolumn{1}{c}{(2)} & \multicolumn{1}{c}{(3)}\\ 
\hline \\[-1.8ex] 
 $\ln(E^{t}_{o,d})$ & 0.594^{***} & 0.757^{***} & 0.767^{***} \\ 
  & (0.031) & (0.024) & (0.027) \\ 
  & & & \\ 
 $\ln(P^{t}_{o,d})$ & 0.331^{***} & 0.307^{***} & 0.356^{***} \\ 
  & (0.100) & (0.086) & (0.103) \\ 
  & & & \\ 
 Constant & 0.136 & -2.229^{**} & -3.087^{**} \\ 
  & (1.148) & (1.028) & (1.206) \\ 
  & & & \\ 
\hline \\[-1.8ex] 
Origin FE & \multicolumn{1}{c}{Y} & \multicolumn{1}{c}{Y} & \multicolumn{1}{c}{Y} \\ 
Destination FE & \multicolumn{1}{c}{Y} & \multicolumn{1}{c}{Y} & \multicolumn{1}{c}{Y} \\ 
\hline \\[-1.8ex] 
Observations & \multicolumn{1}{c}{792} & \multicolumn{1}{c}{828} & \multicolumn{1}{c}{870} \\ 
R$^{2}$ & \multicolumn{1}{c}{0.892} & \multicolumn{1}{c}{0.937} & \multicolumn{1}{c}{0.932} \\ 
Adjusted R$^{2}$ & \multicolumn{1}{c}{0.870} & \multicolumn{1}{c}{0.923} & \multicolumn{1}{c}{0.917} \\ 
Residual Std. Error & \multicolumn{1}{c}{1.778} & \multicolumn{1}{c}{1.364} & \multicolumn{1}{c}{1.510} \\ 
F Statistic & \multicolumn{1}{c}{39.197$^{***}$} & \multicolumn{1}{c}{70.256$^{***}$} & \multicolumn{1}{c}{64.684$^{***}$} \\ 
\hline 
\hline \\[-1.8ex] 
\textit{Note:}  & \multicolumn{3}{r}{$^{*}$p$<$0.1; $^{**}$p$<$0.05; $^{***}$p$<$0.01} \\ 
\end{tabular}
\caption{The relationship between the yearly residuals of the Origin Space expectations and the growth in foreign card expenditures the following year. Mastercard insights.} 
\label{tab:country-space-prediction-2}
\end{table} 

The Origin Space is a tool to provide tourism recommendations. For instance, we can see that tourists from Australia and New Zealand are similar. Colombia has attracted a significant portion of Australian tourists, more in proportion that the Netherlands (although less in absolute levels as shown in Tables \ref{tab:col-origins} and \ref{tab:nld-origins}). The data show New Zealand's share of GDP spent by tourists in Colombia is only around a fourth of Australia's share. Since the two tourist populations are very similar, Colombia's tourism stakeholders might want to assess how to attract more New Zealanders. Another observation is that neither Colombia nor the Netherlands seem to have a significant position in the Asian market. However, Colombia has some opportunities in the Middle East, having some presence from the United Arab Emirates.

The Origin Space is interesting, but we need to assess its predictive power. The first question we ask is if there is a correlation between the observed tourist flows and what one would predict using the correlations of destinations -- the Origin Space edges. Prediction is calculated as follows:

$$ P_{o,d} = \dfrac{\sum \limits_{o' \neq o} (\rho_{o,o'} \times E_{o',d})}{\sum \limits_{o' \neq o} \rho_{o,o'}}, $$

where $\rho_{o,o'}$ is the Origin Space similarity between the two origins $o$ and $o'$, and $E_{o',d}$ is the total amount spent in $d$ by cards issued in $o'$. We then test the simple linear model:

$$ \ln(E_{o,d} + 1) = \alpha + \beta \ln(P_{o,d} + 1) + u_o + u_d + \epsilon_{o,d}, $$

where $u_o$ and $u_d$ are the country of origin and country of destination fixed effects. Both expenditures and our prediction are transformed using the natural logarithm. The correlation is present and significant, with $p < 0.01$.  Table \ref{tab:country-space-prediction-1} reports the coefficient.

The deviation from the regression line in Table \ref{tab:country-space-prediction-1} is predictive of future expenditure levels. The growth prediction regression controls for initial level, and uses our Origin Space estimate:

$$ \ln(E^{t+1}_{o,d} + 1) = \alpha + \beta_1 \ln(E^{t}_{o,d} + 1) + \beta_2 \ln(P^{t}_{o,d} + 1) + u_o + u_d + \epsilon_{o,d},$$

where $E^{t}_{o,d}$ is the expenditures of $o$ cards in $d$ in year $t$ and $P^{t}_{o,d}$ is our expectation given the Origin Space using expenditure data exclusively from year $t$.

Table \ref{tab:country-space-prediction-2} reports the result of this regression. We run the model to predict levels in  2012 (model 1), 2013 (model 2) and 2014 (model 3). $\beta_1$ takes care of the autocorrelation of levels with the previous year. The positive and significant $\beta_2$ -- in all three models -- means that our expectation is indeed significantly associated with changes in foreign card expenditure levels.

Note that in this section we focused on evaluating the expected foreign card expenditures in the country of destination as a whole. However, we could use the same approach to obtain the same expectations at the country-industry level. Such analysis could answer questions like: which industries are currently catering to relatively few tourists and could expand in the future? We leave this analysis as future work.

\subsection{Destination Space}
In this section we perform an operation similar to the one presented in the previous section. In this case, we focus on the destinations instead of on the origins. The creation of the Destination Space follows the same methodology as for the Origin Space: we build a vector of destinations $V_d$ for each destination and we calculate all pairwise correlations. Instead of visualizing the Destination Space as a network -- it would contain too many nodes for a meaningful visualization --, we perform some additional steps. We run a community discovery algorithm on the Destination Space. Community discovery is a popular network problem, the aim of which is to identify functional modules in the network, represented by groups of nodes densely interconnected with each other \cite{coscia2011classification}. We run the Infomap algorithm \cite{rosvall2008maps} to detect our communities, as it is one of the best performing partition algorithms. We then map the resulting destination communities.

\begin{figure}
\centering
\includegraphics[width=.495\columnwidth]{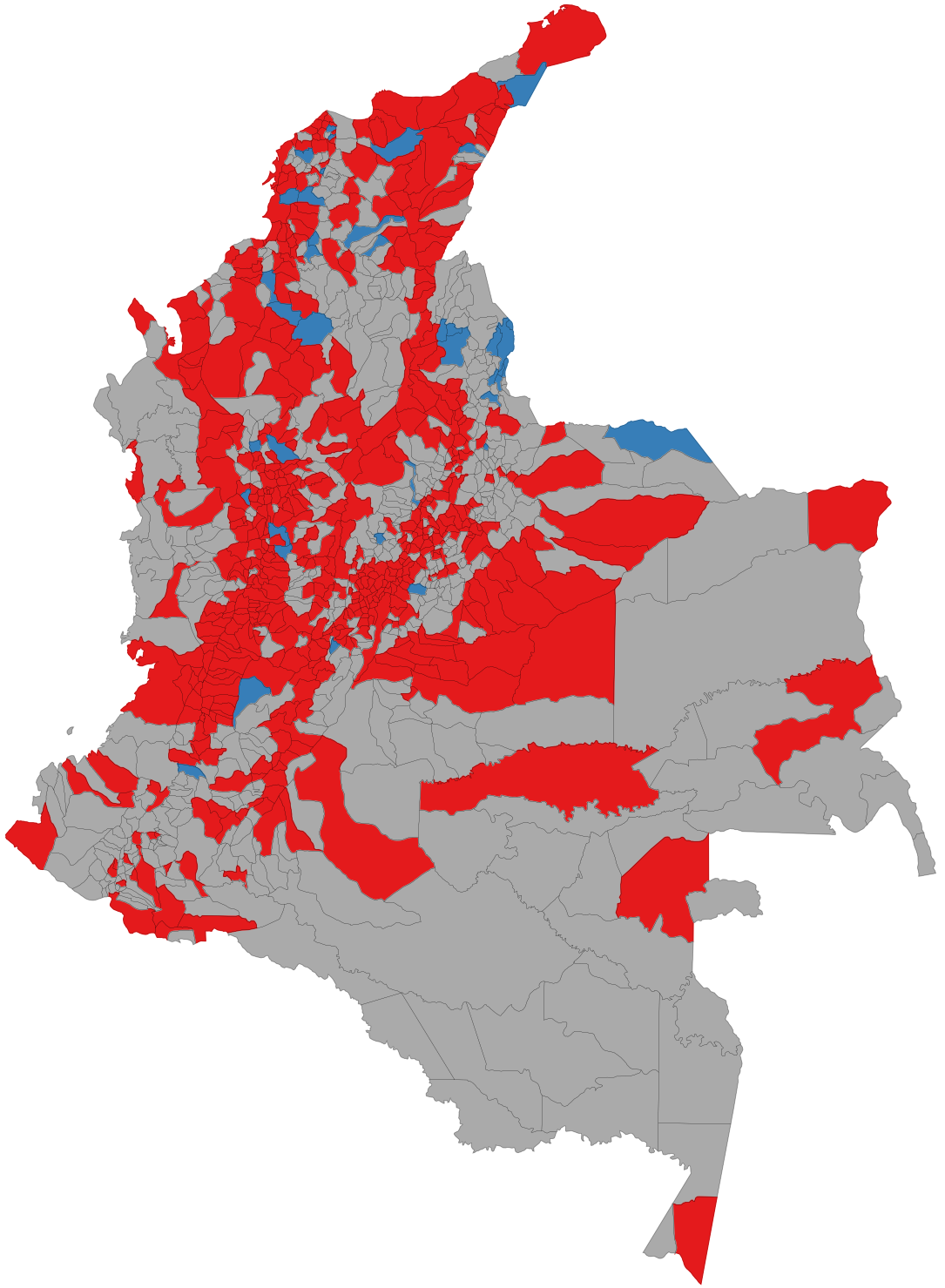}
\includegraphics[width=.495\columnwidth]{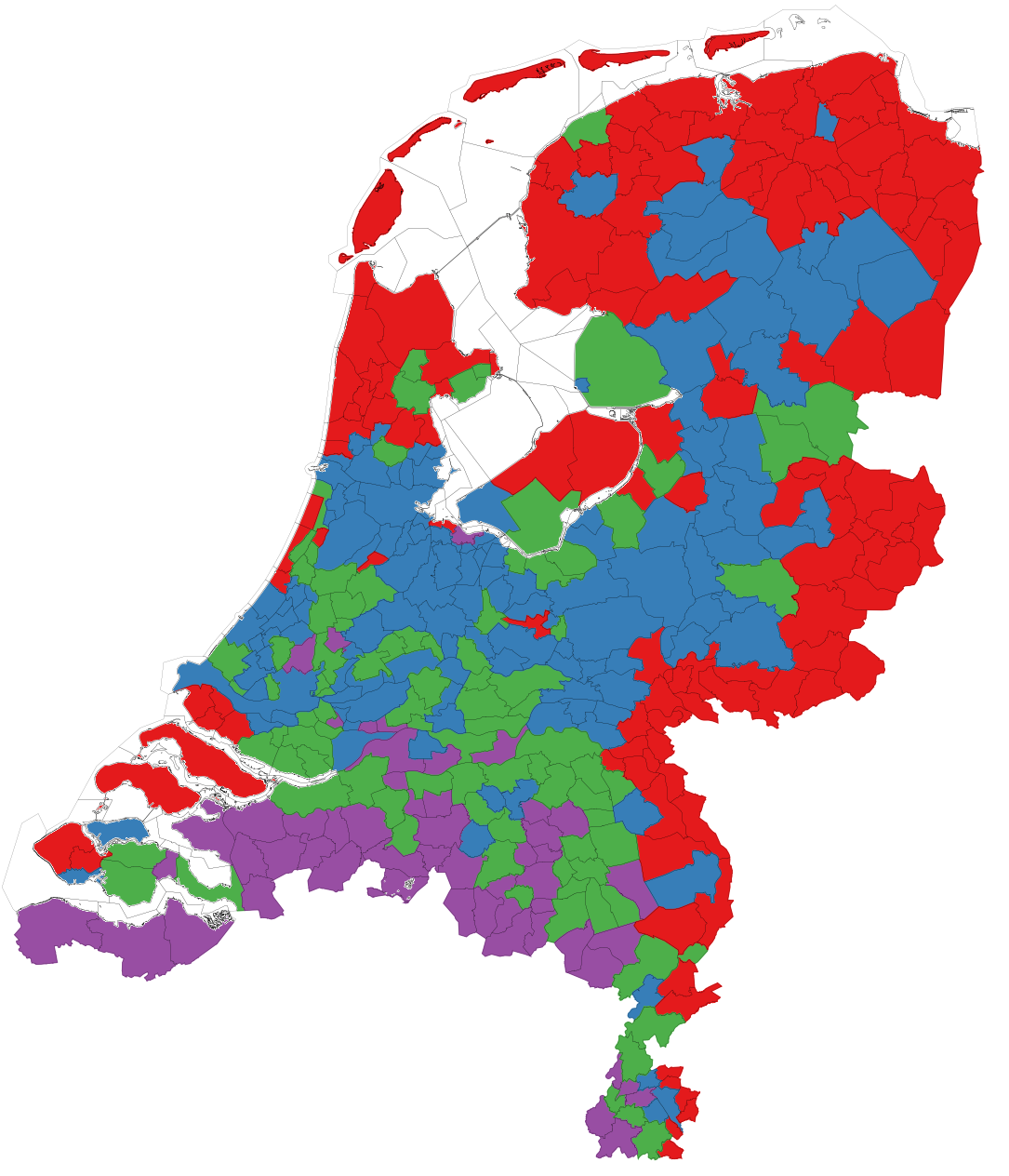}
\caption{The destination clusters according to the portfolio of origins visiting them, for Colombia (left) and the Netherlands (right). Municipalities visited disproportionately by the same countries are coded with the same color. Mastercard insights.}
\label{fig:destination-clusters}
\end{figure}

Figure \ref{fig:destination-clusters} depicts the clusters. Again, we can notice a fundamental difference between Colombia and the Netherlands. In Colombia, the algorithm failed to find meaningful communities. Almost all municipalities for which we analyze spend data are part of one giant central community. Infomap returns a hierarchical community partition, of which here we represent the top level. However, going further down the hierarchy yields tens of clusters, which are hard to visualize and not geographically compact. On the other hand, the Netherlands have a few clear clusters, which are easy to interpret.

\begin{table}[!h]
\centering \footnotesize
\begin{tabular}{l|l|r}
Cluster & Origin & \% of Origin\\
\hline
\hline
\multirow{5}{*}{1} & Poland & 78.84\%\\
 & Cura\c{c}ao & 73.24\%\\
 & Aruba & 73.08\%\\
 & Germany & 71.00\%\\
 & Greece & 66.31\%\\
\hline
\multirow{5}{*}{2} & Venezuela & 97.84\%\\
 & Mexico & 78.99\%\\
 & Japan & 76.21\%\\
 & Brazil & 69.98\%\\
 & Hong Kong & 68.33\%\\
\hline
\multirow{5}{*}{3} & Belgium & 6.68\%\\
 & Poland & 6.55\%\\
 & Germany & 3.51\%\\
 & United Kingdom & 2.76\%\\
 & Switzerland & 2.55\%\\
\hline
\multirow{5}{*}{4} & Belgium & 51.71\%\\
 & France & 5.84\%\\
 & Germany & 4.97\%\\
 & Poland & 3.90\%\\
 & United Kingdom & 3.65\%\\
\end{tabular}
\caption{The characterization of the four destination clusters of the Netherlands. Color map for the clusters in Figure \ref{fig:destination-clusters} (right): 1 = red, 2 = blue, 3 = green, 4 = purple. We report an entry only if it totaled a significant amount of expenditure. Mastercard insights.}
\label{tab:destination-clusters-nld}
\end{table}

Table \ref{tab:destination-clusters-nld} helps with the interpretation of the clusters. We rank origins according to their Origin Relative Expenditure value. We divide the total amount spent on a cluster by the total amount spent in the country. We can characterize each cluster as follows: the red cluster is the East commuting cluster, dominated by Poland and Germany -- its relative expenditure value is not the highest, but the level of expenditure (not reported) is by far the most important --; the blue cluster is the core tourism cluster, dominated by long-range trips; the green cluster is a short-range tourism cluster, dominated by European travelers; and the purple cluster is the South commuting cluster, dominated by Belgium.

The commuting clusters appear to be spending the most, but it is arguably something that municipalities outside those clusters cannot change. A possible way to help municipalities is to understand the core tourism cluster better, so that municipalities in the commuting clusters can also attract tourists. Here, we focus on the attractions that are over expressed in the core-tourism cluster and are absent in the commuting clusters. Many of these attractions are unique: it would be difficult for any municipality in the Netherlands to replicate Amsterdam's historical and cultural heritage. However, we can still investigate the ecosystem around that: the distribution of merchant types.

\begin{table}[!h]
\centering
\begin{tabular}{l|l|r}
Cluster & Origin & \% of Industry\\
\hline
\multirow{5}{*}{1} & Real Estate Services & 64.53\%\\
 & Automotive Fuel & 60.35\%\\
 & Sporting Goods / Apparel / Footwear & 56.44\%\\
 & Construction Services & 50.75\%\\
 & Travel Agencies and Tour Operators & 49.54\%\\
\hline
\multirow{5}{*}{2} & T+E Taxi and Limousine & 95.91\%\\
 & T+E Vehicle Rental & 94.19\%\\
 & Live Performances, Events, Exhibits & 92.45\%\\
 & Health/Beauty/Medical Supplies & 85.75\%\\
 & Jewelry and Giftware & 79.00\%\\
\hline
\multirow{5}{*}{3} & Casino and Gambling Activities & 25.11\%\\
 & Automotive New and Used Car Sales & 15.44\%\\
 & Giftware/Houseware/Card Shops & 12.62\%\\
 & Miscellaneous Personal Services & 11.96\%\\
 & Home Improvement Centers & 9.33\%\\
\hline
\multirow{5}{*}{4} & Grocery Stores & 54.26\%\\
 & Toy Stores & 49.13\%\\
 & Home Improvement Centers & 45.65\%\\
 & Department Stores & 42.76\%\\
 & Wholesale Trade & 42.56\%\\
\end{tabular}
\caption{The characterization of the industries visited per cluster. Color map for the clusters is the same of table \ref{tab:destination-clusters-nld}. We filter out non-statistically significant observations. Mastercard insights.}
\label{tab:destination-clusters-nld-2}
\end{table}

Table \ref{tab:destination-clusters-nld-2} reports the top 5 industries according to their Origin Relative Expenditure value for the clusters. What tourist segments from clusters \#1 and \#4 buy are products it would make little sense to travel a long distance for. On the other hand, the typical merchant types in cluster \#2 are focused on transportation when tourists do not have access to their own vehicle. Tourist segments in cluster \#2 are also interested in live performances, events, exhibits -- as expected -- but also in cosmetics and related products, and jewelry and giftware.

\subsection{Merchant Space}
For the final section of the paper we focus on merchants. Rather than building a fully fledged third space -- after the Origin and Destination spaces -- we limit ourselves to classify each industry as tourism, non-tourism or other. The aims are to quantify the impact on foreign-originated spend in the industries that are not commonly thought as being part of the tourism sector, and to show that these spend categories are not exclusively purchased by commuters, but also in non-negligible quantities by regular tourists.

The starting point is to divide a product in ``tourism'' or ``commuting''. The basis of this classification is aggregated spend data from all six countries including the destination country. We calculate the Pearson and Spearman correlation of each industry with the accommodation industry, on the basis that activities which correlate with hotels are likely to be performed by tourists. We sort industries by their combined correlations. The third of industries with the lowest p-values are classified as ``tourism'' industries. Every product with a p-value equal to or higher than ATMs is a ``commuting'' product. Products not satisfying either constraint are classified as ``other''. The list of industries with their classification is reported in the Appendix.

Figure \ref{fig:industry-split} depicts the distribution of foreign credit/debit card expenditures in these classes. We first focus on the national level distribution in Figure \ref{fig:industry-split} (left). The set of commuting industries are very important sources of foreign expenditures: they represent a quarter of total dollars spent in the Netherlands. In Colombia this percentage is around 30\%.

\begin{figure}
\centering
\includegraphics[width=.495\columnwidth]{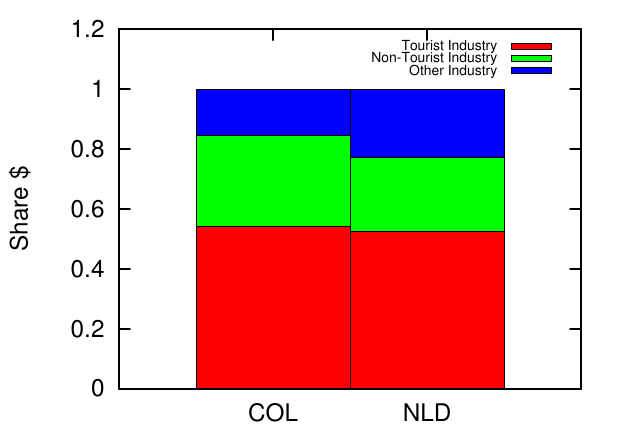}
\includegraphics[width=.495\columnwidth]{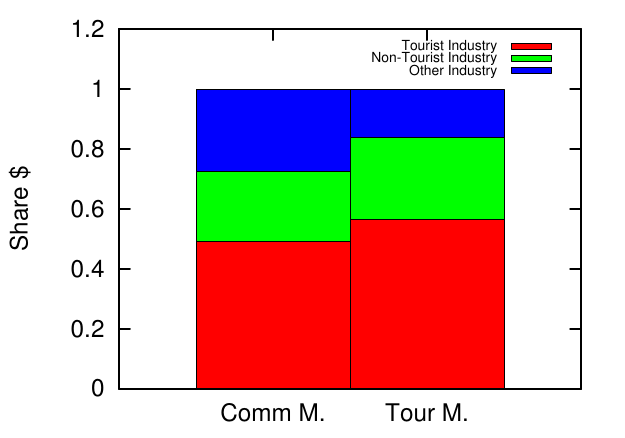}
\caption{The shares of tourism, commuting and other products in different views of foreign card activities. (Left) comparison between Colombia and the Netherlands overall. (Right) Comparison between the Netherlands clusters: tourism (cluster \#2, blue in Figure \ref{fig:destination-clusters} right) and commuting (\#1 and \#4, red and purple in Figure \ref{fig:destination-clusters} right). Mastercard insights.}
\label{fig:industry-split}
\end{figure}

For the Netherlands, we can focus on the subnational level. Here we split expenditures if they happened in cluster \#2 (tourism) or in clusters \#1 and \#4 (commuting) -- see Figure \ref{fig:industry-split} (right). Commuting industries represent a significant share of expenditures in both the commuting and tourism clusters. Tourists spend a non-negligible amount of dollars in industries that are probably not captured by traditional tourism indicators. This amount represents more than 27\% of their expenditures.

\section{Discussion}
In this paper we used unique anonymized and aggregated transaction datasets to examine foreign credit/debit expenditures in a country. We presented an understanding of this data and possible analyses to provide insight as to where tourism is most likely to emerge and the nature of tourist segments' expenditures. In the document we focused particularly on two countries: Colombia and the Netherlands.

We can summarize the results of this descriptive paper in two categories: validation and potential findings. On validation:

\begin{itemize}
\item We expect geographical distance, GDP per capita of the country of origin and other cultural variables to play an important role in determining the attractiveness of a country. We validated this expectation by showing that these three variables account for around 80\% of the variation in tourism expenditures.
\item Municipalities do focus on specific industries to increase their tourism income, and we confirmed that with the anonymized and aggregated transaction data. For instance, we saw that Rotterdam was able to attract more foreign expenditures on cruise tickets than Amsterdam, even if the latter city is a larger tourism hub.
\item Climate plays a detectable role, causing significant differences in the tourism cycle of different countries. Colombia can experience year-round tourism, while the Netherlands attracts most of its foreign visitors during summer. The data allowed us to make this expected distinction.
\end{itemize}

Some of the potential findings of the paper are:

\begin{itemize}
\item Colombia and the Netherlands have a significant difference in the distribution of tourism destinations. Colombia has few centralized attractors, while the Netherlands is more akin to a decentralized self-organizing system.
\item We can build an Origin Space, connecting countries of origin if their tourist segments' expenditures are similar. The Origin Space provides insights as to changes in tourism patterns.
\item We can build a Destination Space, connecting municipalities of destination if they are visited by the same origins. This helps us in classifying both municipalities and industries as proper tourism or commuting destinations.
\item We showed that there is a set of industries that cannot be easily classified as ``tourism'' industries, because their products are usually purchased by local residents -- or commuters. However, the anonymized and aggregated transaction data show that revenues in these industries from actual tourists is not negligible.
\end{itemize}

These findings pave the way to countless research opportunities. These preliminary analyses suggest that a promising research plan can be developed around providing insight as to tourism, its origins, destinations and expenditure pattern, as well as a deeper understanding of how tourism contributes to a country's economy.

\section*{Acknowledgements}
We thank the Mastercard Center for Inclusive Growth for donating anonymized and aggregated transaction data without which this study would not have been possible.

\bibliography{biblio}

\section*{Appendix}

\subsection*{Classification of Industries}

\begin{longtable}{l|l}
Cluster & Industry\\
\hline
\multirow{30}{*}{Tourism} & Accommodations \\
& Bars/Taverns/Nightclubs \\
& Beer/Wine/Liquor Stores \\
& Book Stores \\
& Children's Apparel \\
& Eating Places \\
& Elementary, Middle, High Schools \\
& Family Apparel \\
& Giftware/Houseware/Card Shops \\
& Grocery Stores \\
& Health/Beauty/Medical Supplies \\
& Jewelry and Giftware \\
& Luggage and Leather Stores \\
& Miscellaneous Apparel \\
& Miscellaneous entertainment and recreation \\
& Miscellaneous Vehicle Sales \\
& Newspapers and Magazines \\
& Optical \\
& Other Transportation Services \\
& Photofinishing Services \\
& Photography Services \\
& Specialty Food Stores \\
& T+E Bus \\
& T+E Cruise Lines \\
& T+E Vehicle Rental \\
& Travel Agencies and Tour Operators \\
& Utilities \\
& Variety / General Merchandise Stores \\
& Video and Game Rentals \\
& Women's Apparel \\
\hline
\multirow{49}{*}{Commuting} & Accounting and Legal Services \\
& Advertising Services \\
& Agriculture/Forestry/Fishing/Hunting \\
& Automotive Retail \\
& Automotive Used Only Car Sales \\
& Casino and Gambling Activities \\
& Cleaning and exterminating Services \\
& Clothing, Uniform, Costume Rental \\
& College, University Education \\
& Communications, Telecommunications Equipment \\
& Computer / Software Stores \\
& Consumer Credit reporting \\
& Cosmetics and Beauty Services \\
& Courier Services \\
& Dating Services \\
& Death Care Services \\
& Discount Department Stores \\
& Drug Store Chains \\
& Dry Cleaning, Laundry Services \\
& Employment, Consulting Agencies \\
& Equipment Rental \\
& Financial Services (ATMs) \\
& Florists \\
& Health Care and Social Assistance \\
& Home Furnishings / Furniture \\
& Information Retrieval Services \\
& Insurance \\
& Live Performances, Events, Exhibits \\
& Men's Apparel \\
& Miscellaneous \\
& Miscellaneous Administrative and Waste Disposal Services \\
& Miscellaneous Personal Services \\
& Miscellaneous Professional Services \\
& Miscellaneous Publishing Industries \\
& Miscellaneous Technical Services \\
& Movie and Other Theatrical \\
& Office Supply Chains \\
& Pet Stores \\
& Public Administration \\
& Real Estate Services \\
& Religious, Civic and Professional Organizations \\
& Security, Surveillance Services \\
& Software Production, Network Services and Data Processing \\
& T+E Railroad \\
& T+E Taxi and Limousine \\
& Veterinary Services \\
& Warehouse \\
& Wholesale Clubs \\
& Wholesale Trade \\
\hline
\multirow{20}{*}{Other} & Amusement, Recreation Activities \\
& Arts and Craft Stores \\
& Automotive Fuel \\
& Automotive New and Used Car Sales \\
& Camera/Photography Supplies \\
& Communications, Telecommunications, Cable Services \\
& Construction Services \\
& Consumer Electronics / Appliances \\
& Department Stores \\
& Home Improvement Centers \\
& Maintenance and Repair Services \\
& Manufacturing \\
& Miscellaneous Educational Services \\
& Music and Videos \\
& Professional Sports Teams \\
& Shoe Stores \\
& Sporting Goods / Apparel / Footwear \\
& T+E Airlines \\
& Toy Stores \\
& Vocation, Trade and Business Schools \\
\end{longtable}

\end{document}